\documentclass[usenatbib,fleqn]{mnras}
\usepackage{newtxtext,newtxmath}  
\usepackage{graphicx,hyperref,amsmath,mathrsfs,xspace}
\usepackage[dvipsnames]{xcolor}

\title[Milky Way mass measurement with LMC]{Measuring the Milky Way mass distribution in the presence of the LMC}

\author[Correa Magnus \& Vasiliev]{
Lilia Correa Magnus$^{1}$\thanks{E-mail: lilimagnus@hotmail.es},
Eugene Vasiliev$^{2,3}$\thanks{E-mail: eugvas@lpi.ru}\\
$^1$University of Edinburgh, Peter Guthrie Tait road, Edinburgh, EH9 3FD, UK \\
$^2$Institute of Astronomy, Madingley road, Cambridge, CB3 0HA, UK\\
$^3$Lebedev Physical Institute, Leninsky prospekt 53, Moscow, 119991, Russia
}

\newcommand{\Gaia}{\textit{Gaia}\xspace}
\newcommand{\kms}{km\:s$^{-1}$\xspace}
\newcommand{\masyr}{mas\:yr$^{-1}$\xspace}
\date{Accepted 2021 December 17. Received 2021 December 10; in original form 2021 September 30}

\begin{document}
\label{firstpage}
\pagerange{2610--2630}\volume{511}\pubyear{2022}
\setcounter{page}{2610}
\maketitle

\begin{abstract}
The ongoing interaction between the Milky Way (MW) and its largest satellite -- the Large Magellanic Cloud (LMC) -- creates a significant perturbation in the distribution and kinematics of distant halo stars, globular clusters and satellite galaxies, and leads to biases in MW mass estimates from these tracer populations. We present a method for compensating these perturbations for any choice of MW potential by computing the past trajectory of LMC and MW and then integrating the orbits of tracer objects back in time until the influence of the LMC is negligible, at which point the equilibrium approximation can be used with any standard dynamical modelling approach. We add this orbit-rewinding step to the mass estimation approach based on simultaneous fitting of the potential and the distribution function of tracers, and apply it to two datasets with the latest \Gaia EDR3 measurements of 6d phase-space coordinates: globular clusters and satellite galaxies. We find that models with LMC mass in the range $(1-2) \times 10^{11}\,M_\odot$ better fit the observed distribution of tracers, and measure MW mass within 100~kpc to be $(0.75\pm0.1)\times 10^{12}\,M_\odot$, while neglecting the LMC perturbation increases it by $\sim15$\%.
\end{abstract}


\section{Introduction}   \label{sec:intro}

The study of stellar dynamics in the Milky Way (MW) and its neighbourhood flourishes in recent years thanks to the vast amount of observational data provided by the \Gaia satellite and numerous ground-based spectroscopic surveys. One aspect of this analysis is the measurement of the mass distribution up to the virial radius%
\footnote{here $r_\text{vir}$ is defined as the radius within which the mean density is 102 times higher than the cosmic density of matter, and is related to the virial mass as $r_\text{vir} = (M_\text{vir}/10^{12}\,M_\odot)^{1/3} \times 260$~kpc.}
of the MW, primarily based on the kinematics of various tracer populations: distant halo stars, globular clusters and satellite galaxies. Commonly used methods rely on the Jeans equations or on simultaneous modelling of the tracer distribution function (DF) and the MW gravitational potential with simple analytic expressions, such as the power-law mass estimator of \citet{Watkins2010}, variants of which have been applied to globular clusters or satellites by \citet{Eadie2019} and \citet{Fritz2020}, respectively. More complicated DF families in action space were employed by \citet{Posti2019} and \citet{Vasiliev2019b} for the globular cluster population, all based on the previous \Gaia data release (DR2). The most recent Early data release 3 \citep{Brown2021} improved the accuracy of proper motions (PM) by a factor of two or better, and the new measurements of cluster and satellite PM by \citet{Vasiliev2021a}, \citet{McConnachie2020}, \citet{Li2021}, \citet{Battaglia2021} are awaiting application to the MW mass modelling.

At the same time, a new challenge has emerged in dynamical analysis of the MW outskirts: the Large Magellanic Cloud (LMC), which just passed the pericentre of its orbit, appears to be a massive enough satellite to cause a significant perturbation in the motion of stars and other objects at distances beyond a few tens kpc. Various recent estimates of the LMC mass put it in the range $(1-2)\times 10^{11}\,M_\odot$ (see \citealt{Shipp2021} and references therein), i.e., only $5-10$ times smaller than the MW itself.

The effects of the LMC on the MW system are manifold. First, it has brought a host of its own satellites with it, some of which are now stripped and became MW satellites, as discussed, e.g., in \citet{Battaglia2021}. Second, it deflects MW stars, satellites and stellar streams that pass in its vicinity, sometimes dramatically changing their orbits (e.g., the southern portion of the Orphan--Chenab stream, \citealt{Erkal2019}). The MW halo stars deflected by the LMC form a density wake behind its orbit, as per the classical dynamical friction scenario, which might have been detected as the Pisces overdensity \citep{Belokurov2019}. However, another equally important and perhaps less obvious effect is caused by the global perturbation induced by the LMC on the outer parts of the MW, which arises from a combination of two factors. On the one hand, the LMC and the MW move under mutual gravitational forces as a binary system, so the reflex motion of the MW produces an acceleration of the associated Galactocentric reference frame, as first stressed by \citet{Gomez2015}. The non-inertial frame is not necessarily a concern: the fact that the entire MW is also pulled towards the Andromeda galaxy and towards the Virgo cluster has little effect on the motion of test bodies in the Galaxy because the acceleration is spatially uniform, and the Galactocentric reference frame is free-falling, as in the Einstein's elevator \textit{gedankenexperiment}. But given the proximity of the LMC, distant objects in the MW halo feel a different acceleration from it than the MW centre -- in other words, the MW becomes deformed by the LMC (and, of course, vice versa). One could view this from a different perspective: the dynamical time in the inner Galaxy (e.g., in the Solar neighbourhood) is short enough that the LMC passage is an adiabatic perturbation for these objects, and their Galactic orbits are little affected. But the outer halo objects are not able to adjust their orbits rapidly enough and thus are displaced w.r.t.\ the MW centre -- or, better to say, the central part of the MW rapidly swings towards the LMC on its pericentre passage, while its outer parts largely stay put. In the end, the resulting dipole perturbation is manifested both in the density and in the kinematics of the outer MW halo \citep[e.g.,][]{Cunningham2020, GaravitoCamargo2020}, and both effects have been detected recently: the former in \citet{Conroy2021}, the latter in \citet{Erkal2021} and \citet{Petersen2021}.

Naturally, these perturbations invalidate the assumption of dynamical equilibrium that lies at the heart of most dynamical modelling methods. \citet{Erkal2020b}, using a simple power-law mass estimator from \citet{Watkins2010}, concluded that the neglect of this perturbation causes an upward bias in the inferred MW mass by up to 50\%, depending on the LMC mass and the distance within which the MW mass is measured. More recently, \citet{Deason2021} measured the MW mass out to 100~kpc from the kinematics of halo stars, while approximately correcting for the LMC-induced bias by shifting the stellar velocities by a distance-dependent offset calibrated in $N$-body simulations, although this changed the resulting mass estimate by $\lesssim 2\%$.

In this work, we take a step further and develop an orbit-rewinding method for compensating the perturbation from the LMC in any given MW potential model, which can be used with any classical dynamical modelling approach based on the equilibrium assumption. In subsequent analysis, we employ the method for measuring the gravitational potential of the MW by fitting a parametric DF to the population of tracer objects, and apply it to two classes of tracers: globular clusters and satellite galaxies, using the latest \Gaia EDR3 measurements. Section~\ref{sec:method} describes our modelling method, including the orbit-rewinding step and another novel aspect -- the treatment of possible outliers in the kinematic sample.  Section~\ref{sec:tests} demonstrates the performance of the method on mock datasets. Section~\ref{sec:MW} specifies the model setup for the actual MW and the observational catalogues that we use in the analysis, and Section~\ref{sec:results} presents the results of our fits: the MW mass profile, the properties of the tracer populations (primarily satellite galaxies), and discusses the role of the LMC in their kinematics. Section~\ref{sec:summary} wraps up.

\section{Method}   \label{sec:method}

Our approach falls within the class of likelihood-based forward modelling methods. For any choice of parameters describing the DF of the tracer population, the (initial, unperturbed) potential of the MW, and the LMC mass, we evaluate the likelihood of the observed dataset in this model, taking into account the observational errors and the perturbation from the LMC. We then explore the parameter space with the Monte Carlo Markov Chain method. Below we describe these steps in more detail. 

Our analysis pipeline relies on the \textsc{Agama} stellar-dynamical framework \citep{Vasiliev2019a}, which provides a wide choice of gravitational potentials, DFs, orbit integration and other tools, but the general idea can be applied in a broader context and with other modelling platforms.

\subsection{Orbit rewinding}  \label{sec:orbit_rewinding}

The key novel aspect of our work is the orbit rewinding procedure for compensating the LMC perturbation. It is applied to the tracer population for each choice of model parameters, and consists of two steps: first we reconstruct the past trajectories of the MW and the LMC under mutual gravitational forces, and then integrate the orbits of tracers backward in time up to the point when the LMC perturbation was negligible.

Although the LMC was usually considered as just one of many MW satellites, whose orbit is determined entirely by the gravitational potential of the Galaxy and the dynamical friction, the relatively large mass ratio (likely between $1:10$ and $1:5$) between the two galaxies makes their interaction far more complex. Both galaxies exert gravitational force onto each other, but are also deforming in the process of interaction, so the forces acting on the central parts of both galaxies have additional contribution from their own distorted mass distribution in the outer parts. As discussed in \citet{Vasiliev2021c}, the ``self-gravity'' is actually the dominant mechanism in the evolution of orbital angular momentum: the commonly used approximation based on the Chandrasekhar dynamical friction formula produces qualitatively different evolution than found in $N$-body simulations. Nevertheless, these nonlinear phenomena kick in largely after the first pericentre passage. In the case of the LMC, we are fortunate that it just passed its pericentre very recently, likely for the first time \citep[e.g.,][]{Kallivayalil2013}, and therefore its past trajectory might be reasonably well approximated by a simple kinematic model with two rigid mutually gravitating galaxies, without worrying too much about their internal deformations. This approach adequately captures the most important effect of the MW reflex motion, as highlighted by \citet{Gomez2015}, and was used in a number of subsequent papers (e.g., \citealt{Jethwa2016,Erkal2019, Vasiliev2021b, Shipp2021}); here we briefly summarize it.

The positions and velocities of the MW and the LMC evolve according to the coupled ODE system:
\begin{equation}  \label{eq:lmc_mw_orbit}
\begin{aligned}
\dot{\boldsymbol{x}}_\text{MW} &= \boldsymbol{v}_\text{MW}, \\
\dot{\boldsymbol{v}}_\text{MW} &= -\nabla\Phi_\text{LMC}(\boldsymbol{x}_\text{MW}-\boldsymbol{x}_\text{LMC}), \\
\dot{\boldsymbol{x}}_\text{LMC} &= \boldsymbol{v}_\text{LMC}, \\
\dot{\boldsymbol{v}}_\text{LMC} &= -\nabla\Phi_\text{MW}(\boldsymbol{x}_\text{LMC}-\boldsymbol{x}_\text{MW}) + \boldsymbol{a}_\text{DF}, 
\end{aligned}
\end{equation}
where $\Phi_\text{MW}$ and $\Phi_\text{LMC}$ are static, non-deforming potentials of both galaxies, and $\boldsymbol{a}_\text{DF}$ is the Chandrasekhar dynamical friction acceleration \citep[Equation~8.6]{Binney2008}:
\begin{equation}  \label{eq:dynfric}
\begin{aligned}
\boldsymbol{a}_\text{DF} &\equiv \frac{-4\pi\, \rho_\text{MW}\, G^2\, M_\text{LMC} \ln \Lambda}{v^2}\left[\text{erf}(X) - \frac{2X\,\exp(-X^2)}{\sqrt{\pi}}\right] \frac{\boldsymbol{v}}{v} , \\
X &\equiv v/ \sqrt{2}\sigma_\text{MW} .
\end{aligned}
\end{equation}
Here $\rho_\text{MW}$ and $\sigma_\text{MW}$ are the density and velocity dispersion of the host galaxy at the given position%
\footnote{The velocity dispersion could be obtained from the Jeans equation for each choice of MW potential, but for simplicity, we adopt a universal profile $\sigma_\text{MW} = 150 / \big(1 + |\boldsymbol x_\text{LMC}-\boldsymbol x_\text{MW}| / 100\,\text{kpc}\big)$~\kms, having checked that the results are insensitive to it.}%
, $M_\text{LMC}$ is the LMC mass, $\ln\Lambda$ is the Coulomb logarithm, which we discuss further in Section~\ref{sec:test_lmc_orbit}. In other words, the centres of both galaxies move as point masses in each other's gravitational potential, but these potentials are rigidly attached to the moving points. Since the interaction is manifestly non-symmetric, these equations do not conserve any of the classical integrals of motion, but nevertheless approximate the actual trajectories reasonably well, as we demonstrate in the later section. The equations are integrated back in time from the present-day position and velocity of the LMC in the Galactocentric frame for a time $T_\text{rewind}$. The choice of this rewinding time is somewhat arbitrary, but the LMC needs to recede sufficiently far away so that its perturbation is negligible. We find that for $T_\text{rewind}=2$~Gyr, the LMC moves from its present-day distance of $\sim 50$~kpc to $200-300$~kpc, and even for the most massive MW potentials, still remains near the apocentre of its orbit (for lower-mass MW, the orbit is actually unbound).

In the second step, the orbits of all tracer objects are integrated backward in time for the same interval $T_\text{rewind}$ in the combined MW+LMC potential. Of course, the MW centre as computed from Equation~(\ref{eq:lmc_mw_orbit}) moves away from origin, but it is more convenient to carry out subsequent steps in the non-inertial reference frame pinned to the MW centre, so we initialize a composite time-dependent potential of the two galaxies as follows. The MW potential is fixed at origin, the LMC is represented by a rigid analytic potential moving along the pre-computed trajectory $\boldsymbol{x}_\text{LMC}(t)-\boldsymbol{x}_\text{MW}(t)$, and we add a spatially uniform but time-dependent acceleration $\boldsymbol{a}_\text{non-inertial}(t) \equiv -\ddot{\boldsymbol{x}}_\text{MW}(t)$. The positions and velocities of tracers at the moment $T_\text{rewind}$ in the past are then assumed to represent the original, unperturbed state of the MW before the LMC arrival. If desired, we may again integrate the orbits from these past coordinates forward up to present time in a static potential of the MW alone, obtaining the ``compensated'' present-day coordinates that the objects would have had if the LMC did not exist. However, since the past and the compensated coordinates belong to the same orbit and differ only in phase angles, they are equivalent for the purpose of dynamical modelling, which uses only the integrals of motion of tracer objects. As discussed in Section~\ref{sec:test_orbit_rewinding}, orbits in the inner part of the Galaxy are almost unperturbed by the LMC, so we may save a considerable effort by not rewinding them at all.

\subsection{Distribution function fitting approach} \label{sec:df_fit}

The assumption of dynamical equilibrium underpins almost all classical methods for measuring the gravitational potential. According to the Jeans theorem \citep[Chapter~4.2 in][]{Binney2008}, in the this case the DF $f$ of any tracer population may depend only on the integrals of motion $\mathcal I(\boldsymbol x, \boldsymbol v ;\; \Phi)$ in the given potential $\Phi$ (the latter represents the total mass distribution and needs not be related to the density profile of the tracers). Although many studies prefer to constrain the gravitational potential from the DF moments using the Jeans equations, in this work we choose to work with the DF itself. Namely, we optimize the parameters $\boldsymbol{\Pi}$ of the potential $\Phi(\boldsymbol{x};\;\boldsymbol\Pi)$ and the DF $f(\mathcal{I};\;\boldsymbol\Pi)$ simultaneously to maximize the log-likelihood of drawing the observed 6d phase-space coordinates $\boldsymbol w \equiv \{\boldsymbol x, \boldsymbol v\}$ of tracer objects in the catalogue:
\begin{equation}  \label{eq:likelihood}
\ln\mathcal{L} = \sum_{i=1}^{N_\text{tracers}} \ln f\big( \mathcal I(\boldsymbol w_i; \; \Phi )\big).
\end{equation}

The DF-fitting approach has been shown to perform well in a variety of contexts, even with missing data dimensions, as explored e.g.\ by \citet{Read2021} on mock datasets representing stars in dwarf galaxies with resolved stellar kinematics. In practice, we need to choose a suitable family of parametric DFs and potentials, which are detailed below.

A flexible model for the density profile of spherical or ellipsoidal stellar systems is a double-power-law \citet{Zhao1996} profile, augmented with an optional exponential cutoff (\texttt{Spheroid} model in \textsc{Agama}):
\begin{equation}  \label{eq:spheroid_density}
\begin{aligned}
\rho &= \rho_0 \left (\frac{\tilde r}{r_\text{scale}} \right)^{-\gamma}
\left [1+\left (\frac{\tilde r}{r_\text{scale}} \right)^{\alpha} \right]^{\frac{\gamma - \beta}{\alpha}} \!\!
\exp\left[-\left (\frac{\tilde r}{r_\text{cut}}\right)^\xi \right] , \\
\tilde r &\equiv \sqrt{R^2+(z/q)^2}\quad \mbox{(ellipsoidal radius)} ,
\end{aligned}
\end{equation}
where $r_\text{scale}$ is the scale radius, $\gamma$ is the asymptotic inner slope $-\mathrm{d}\ln\rho/\mathrm{d}\ln r$, $\beta$ is the outer slope, $\alpha$ is the steepness of transition between the two regimes, $r_\text{cut}$ is the optional outer cutoff radius, and $\xi$ is the cutoff strength. We use this profile for the MW dark halo, the LMC (represented by a spherical NFW profile, i.e., $\gamma=1, \beta=3, \alpha=1$), and for the density profile of tracers in some cases. In particular, our primary model for the MW dark halo has no cutoff and 5 free parameters ($\rho_0$, $r_\text{scale}$, $\gamma$, $\beta$, $\alpha$), but we additionally consider the \citet{Einasto1965} profile with 3 free parameters ($\rho_0$, $r_\text{cut}$, $\xi$, setting $\beta=\gamma=0$); both Zhao and Einasto models are special cases of the \texttt{Spheroid} model, but the full 7-parameter model has too much flexibility to be practical. The baryonic component of the MW is represented by an exponential disc model and a spheroid bulge, as described in \citet{McMillan2017}.

For the tracer DF, we have two alternative choices. The first one is the \texttt{QuasiSpherical} DF constructed from the given tracer density $\rho_\text{tr}$ in the given total potential $\Phi$, with two tunable parameters specifying the velocity anisotropy: the value of anisotropy coefficient $\beta_0 \equiv 1 - \sigma_\text{tan}^2 / 2\sigma_\text{rad}^2$ at small radii, and the anisotropy radius $r_\text{a}$, beyond which the distribution gradually becomes dominated by radial orbits. Its functional form is $f(E,L) = L^{-2\beta_0}\,f_Q\big(E + L^2/2r_\text{a}^2\big)$, with the function $f_Q$ constructed numerically from the provided $\rho_\text{tr}$ and $\Phi$ using the generalization of the Eddington inversion formula by \citet{Cuddeford1991}. For some combinations of parameters, this integral inversion may result in an unphysical DF attaining negative values for some $E,L$, in which case these parameters are discarded from further consideration. 
This model can be used even with non-spherical potentials, but is limited to systems whose equidensity contours match the equipotential contours and without net rotation. 

A more flexible alternative is offered by DFs in the space of actions $\boldsymbol{J} \equiv \{J_r,\, J_\phi,\, J_z\}$ (Section~3.5 in \citealt{Binney2008}), which are a convenient choice for integrals of motion $\mathcal I$, and can be reasonably accurately computed from $\{\boldsymbol x, \boldsymbol v\}$ in any axisymmetric potential, using the St\"ackel fudge method \citep{Binney2012}. We use the following DF family with 9 free parameters, which is adapted from \citet{Posti2015}:
\begin{equation}  \label{eq:DPL_DF}
\begin{aligned}
f(\boldsymbol{J}) &= f_0\;
\left[ 1 + \left(\frac{J_0}{h(\boldsymbol{J})}\right)^\eta \right]^{\frac{\Gamma}{\eta}}
\left[ 1 + \left(\frac{g(\boldsymbol{J})}{J_0}\right)^\eta \right]^{\frac{\Gamma-\mathrm{B}}{\eta}} \\
 &\times \left[1 + \tanh\frac{\varkappa J_\phi}{J_r+J_z+|J_\phi|}\right] , \\
g(\boldsymbol{J}) &\equiv g_r J_r + g_z J_z\, + (3-g_r-g_z)\, |J_\phi|, \\
h(\boldsymbol{J}) &\equiv h_r J_r + h_z J_z   + (3-h_r-h_z)   |J_\phi|.
\end{aligned}   
\end{equation}
Here the dimensionless coefficients $\Gamma$ and B are related to the density slope at small and large radii, respectively, $\eta$ is the steepness of transition between the two asymptotic regimes,  $J_0$ sets the dimensional scale action, $\varkappa$ defines the amount of streaming motion (rotation), and the mixing coefficients $g_{r,z}$ and $h_{r,z}$ control the velocity anisotropy and spatial flattening at large and small radii, respectively. The normalization $f_0$ is determined from the condition that the total mass (the integral of the DF over the 3d action space) is unity. This DF is very similar to the built-in \texttt{DoublePowerLaw} DF in \textsc{Agama}, with an important difference regarding the rotation: the parameter $\varkappa$ controls the ratio of mean azimuthal velocity to its dispersion at all radii, instead of large radii only, as in the original definition; this better matches the kinematics of observed tracer populations in our study. The density generated by this DF depends on its parameters and on the potential in a rather non-trivial way, but we do not need to know it explicitly: the likelihood of the model is computed from the 6d phase-space coordinates translated into action space, not from the 3d density profile of tracers. In practice, it is flexible enough to represent approximately the two-power-law \texttt{Spheroid} profile and to have shape and velocity anisotropy adjustable separately at small and large radii.

\subsection{Treatment of observational errors}  \label{sec:obs_errors}

Any observational data come with associated uncertainties, which should be taken into account in model fitting. The likelihood of the given datapoint in the model is given by the convolution of the model DF with the error distribution of this point:
\begin{equation}  \label{eq:error_convolution_integral}
f\big( \boldsymbol w^\text{(obs)}_i \big) =
\int f\big( \boldsymbol w^\text{(true)} \big)\, E\big( \boldsymbol w^\text{(obs)}_i \,|\, \boldsymbol w^\text{(true)}, \epsilon_i \big)\; \mathrm{d}\boldsymbol w^\text{(true)} ,
\end{equation}
where $E$ is the normalized probability of measuring $w^\text{(obs)}_i$ given the true phase-space coordinates $\boldsymbol w^\text{(true)}$ and observational uncertainties $\epsilon_i$. In practice, the uncertainties on distance, line-of-sight velocity and PM can be assumed to follow independent normal distributions with the standard deviation (or the full $2\times2$ covariance matrix for the two PM components) quoted in the catalogue. For our sample of tracers (globular clusters and satellite galaxies), the relative uncertainty in the distance is typically at the level of a few percent, in velocity -- a few \kms, and in PM -- $0.02-0.1$~\masyr, corresponding to the transverse velocity error of $10-50$~\kms at 100~kpc (sometimes with a significant correlation between the two PM components, which is fully taken into account). We evaluate the above four-dimensional convolution integral with the Monte Carlo method, replacing it with a sum over $N_\text{samp}$ random samples from the joint error distribution of measured values translated into the Galactocentric position and velocity:
\begin{equation}  \label{eq:error_convolution_sum}
f\big( \boldsymbol w^\text{(obs)}_i \big) \approx \frac{1}{N_{\text{samp},i}}
\sum_{s=1}^{N_{\text{samp},i}} f\big( \boldsymbol w^{(s)}_i \big).
\end{equation}
To reduce the impact of the Poisson error in Monte Carlo integration, we use the same set of random samples for all model evaluations, following \citet{McMillan2013}. This approach can be used even in the case that some dimensions of data, such as line-of-sight velocity, are missing: in this case, we need to integrate over a flat distribution of possible velocities in a wide enough range (greater than the escape velocity). However, if the model DF varies widely across this range, the Monte Carlo sum will be dominated by only a few sampling points, thus increasing the discreteness noise; to mitigate this, a more sophisticated importance sampling strategy could be designed (e.g., as described in Section~3.4 of \citealt{Read2021} and Section~4.3.2 of \citealt{Hattori2021}). In our case, we restrict the input dataset to objects with all 6d phase-space coordinates known, and achieve sufficient accuracy with $N_\text{samp}=100$ for globular clusters and $N_\text{samp}=1000$ for satellites.

\subsection{Treatment of outliers}  \label{sec:outliers}

The modelling approach based on the Jeans theorem assumes that the tracer population is fully virialized; however, this might not be true for all Galactic satellites. Even a single non-virialized object with a significantly higher-than-average energy, such as Leo~I, may drive the mass estimate up by a few tens percent \citep[e.g.,][]{Watkins2010}. To mitigate this complication, we allow for a possibility that some objects may not belong to the main population, but instead are described by another DF of the ``unmixed'' population, which is an umbrella term for objects that are unbound, infalling for the first time, or have passed pericentre once but are on orbits with apocentres beyond the virial radius (``splashback'' orbits). Based on a large number of cosmological simulations, \citet{Li2020b} found that the velocities of infalling satellites at the moment of crossing the virial radius of the host halo are well described by a log-normal distribution:
\begin{equation}  \label{eq:velocity_distribution_unmixed}
p(v)\,\text{d}v = \frac{1}{\sqrt{2\pi}\,\sigma_\text{un}} \exp\left[-\frac{\ln^2(v/v_\text{un})}{2\sigma_\text{un}^2}\right] \frac{\text{d}v}{v},
\end{equation}
with $v_\text{un} = 1.2 v_\text{circ}(r_\text{vir}) = 1.2\sqrt{G M_\text{vir}/r_\text{vir}}$ and $\sigma_\text{un} = 0.2$ (dimensionless). The above expression is for the velocity magnitude, and the DF in the 6d phase space is $\displaystyle f(\boldsymbol{x},\boldsymbol{v})\big|_{|\boldsymbol x|=r_\text{vir}} = \rho_\text{un}(r_\text{vir})\, p\big(|\boldsymbol v|\big) \big/ (4\pi v^2)$, where $\rho_\text{un}(r_\text{vir})$ is the density of infalling objects at virial radius. Under the approximation of a stationary host galaxy potential, we may use the Jeans theorem to express this DF as a function of energy (an integral of motion) and evaluate it anywhere in space, not only at the virial radius:
\begin{equation}  \label{eq:df_unmixed}
f_\text{un}(E) = \frac{\displaystyle\rho_\text{un}(r_\text{vir})\, \exp\left[
-\frac{\ln^2\Big\{2\big[E-\Phi(r_\text{vir})\big]/v_\text{un}^2\Big\}}
{8\sigma_\text{un}^2}\right]}
{16\pi^{3/2}\,\sigma_\text{un}\,\big[E-\Phi(r_\text{vir})\big]^{3/2}} .
\end{equation}
Note that the DF is still defined as a probability distribution in the 6d phase space, not in energy space. For instance, the density generated by it at a radius $r$ is computed as
\begin{equation}
\begin{aligned}  \label{eq:density_unmixed}
\rho_\text{un}(r) &= \iiint f_\text{un} \big(E=\Phi(r)+v^2/2\big)\;\text{d}^3v \\
&= \int_0^\infty 4\pi\, v^2\,f_\text{un}\big(\Phi(r)+v^2/2\big)\;\text{d}v \\
&= \int_{\Phi(r_\text{vir})}^\infty 4\pi\,\sqrt{2\big[E-\Phi(r)\big]}\; f_\text{un}(E)\;\text{d}E .
\end{aligned}
\end{equation}
The density increases rather slowly towards small radii at $r<r_\text{vir}$ and decreases more rapidly at larger radii, but the integral of $\rho_\text{un}(r)$ over the entire space is infinite -- this is not unexpected, since the velocity distribution contains a tail of objects with $E>0$ that can travel anywhere. However, since our tracer population has limited extent, we only need to normalize $\rho_\text{un}$ to a unit total mass within some fiducial radius $r_\text{out}$, which we fix to 300~kpc (the most remote object in our sample is Leo~I at a Galactocentric distance $262\pm12$~kpc). For each choice of the MW potential, we compute the virial mass and radius, then evaluate the normalization factor $\rho_\text{un}(r_\text{vir})$ that produces a unit mass within $r_\text{out}$; this fully defines the DF of outliers. A suitable approximation for our range $0.8\le r_\text{vir}/r_\text{out}\le 1.2$ is $\rho_\text{un}(r_\text{vir}) = 0.81\,\frac{3}{4\pi\,r_\text{out}^3}\,(r_\text{vir}/r_\text{out})^{-1/2}$ almost independently of the potential. Note that we do not explicitly need $\rho_\text{un}(r)$ to compute the likelihood of an object in the model: it is given by $f_\text{un}(\boldsymbol x,\, \boldsymbol{v})$, which depends only on its energy in the given potential.

The possibly unmixed outliers are thus treated in a standard mixture model approach. We assume that any object may belong to either of the two populations described by their unit-normalized DFs: $f_\text{bound}\big(\boldsymbol J(\boldsymbol x,\,\boldsymbol v) \big)$ is the action-space DF of bound satellites, and $f_\text{un}\big(E(\boldsymbol x,\,\boldsymbol v)\big)$ is the phase-space DF of interlopers. Let $\eta$ be the overall fraction of the first population in the sample. The mixture DF 
\begin{equation}  \label{eq:df_mixture}
f_\text{mix} = \eta\, f_\text{bound} + (1-\eta)\, f_\text{un},
\end{equation}
evaluated at the phase-space coordinates of each object, now gives the likelihood of this object in the model. The posterior probability of belonging to the first population for $i$-th object is 
\begin{equation}  \label{eq:posterior_outlier_probability}
\eta_i = \frac{\eta\, f_\text{bound}(\boldsymbol x_i,\,\boldsymbol v_i)}
{\eta\, f_\text{bound}(\boldsymbol x_i,\,\boldsymbol v_i) + (1-\eta)\, f_\text{un}(\boldsymbol x_i,\,\boldsymbol v_i)},
\end{equation}
and although we have not specified the overall fraction $\eta$ in advance, it is easy to see that the likelihood is maximized when $\sum_{i=1}^{N_\text{obj}} \eta_i = N_\text{obj}\, \eta$. Indeed, taking the derivative of the total log-likelihood w.r.t.\ $\eta$ and equating it to zero, we find
\begin{equation}
\begin{aligned}
0 &= \frac{\partial}{\partial \eta} \sum\nolimits_{i=1}^{N_\text{obj}} \ln \big[ \eta\, f_{\text{bound},i} + (1-\eta)\, f_{\text{un},i} \big] \\
&= \sum\nolimits_{i=1}^{N_\text{obj}}  \frac{f_{\text{bound},i} - f_{\text{un},i}}{\eta\, f_{\text{bound},i} + (1-\eta)\, f_{\text{un},i}} \\
&= \frac{1}{\eta-1} \sum\nolimits_{i=1}^{N_\text{obj}} \frac{(\eta-1)\,f_{\text{bound},i} + (1-\eta)\,f_{\text{un},i}}{\eta\, f_{\text{bound},i} + (1-\eta)\, f_{\text{un},i}} \\
&= \frac{1}{\eta-1} \left[ N_\text{obj} - \sum\nolimits_{i=1}^{N_\text{obj}} \frac{f_{\text{bound},i}}{\eta\, f_{\text{bound},i} + (1-\eta)\, f_{\text{un},i}} \right] \\
&= \frac{1}{\eta(\eta-1)} \left[ \eta N_\text{obj} - \sum\nolimits_{i=1}^{N_\text{obj}} \eta_i \right].
\end{aligned}
\end{equation}
This simple equation for the auxiliary parameter $\eta$ is easy to solve for each choice of parameters defining the two DFs, thus $\eta$ is not a free parameter and does not increase the complexity of the model (recall that $f_\text{un}$ is fully specified by the host potential).

The addition of this extra ingredient in the model stabilizes it against outliers at very little extra cost. Of course, its performance depends on the validity of the assumption that outliers are reasonably well described by the DF (Equation~\ref{eq:df_unmixed}), but in practice it seems adequate. As a side benefit, the posterior probability of being an outlier (Equation~\ref{eq:posterior_outlier_probability}) can be computed for all objects in our sample and plotted against other model parameters (e.g., virial mass). We use the mixture DF only for the satellite population, since all globular clusters have safely negative energies for any reasonable potential.

\subsection{Monte Carlo simulations}  \label{sec:mcmc}

The above sections describe the steps for constructing a single dynamical model and evaluating its likelihood against the chosen tracer dataset(s). We then explore the parameter space of models with the Markov Chain Monte Carlo (MCMC) approach, using the \textsc{emcee} package \citep{ForemanMackey2013}. We evolve 50 walkers for a few thousand steps, monitoring the convergence by analyzing the evolution of parameters and the overall distribution of likelihoods in the chain, and take the last 500 steps to construct the posterior distribution of model parameters. In the most complete setup (two tracer populations, marginalizing over observational uncertainties and performing orbit rewinding for each choice of model parameters), the evaluation of a single model takes less than a second on a 32-core workstation, and a few times faster without rewinding. Thus the entire analysis pipeline runs in less than a day of wall-clock time.

\section{Tests on mock datasets}  \label{sec:tests}

We test the performance of the method on several mock datasets constructed as follows.

\subsection{Variants of mock models}  \label{sec:mock_variants}

First we choose the structural parameters of the MW, using either a simplified spherical isotropic model with a mass profile closely resembling a realistic galaxy, or more complicated models containing a stellar disc and a dark halo; the latter could be spherical or non-spherical. The LMC is always represented by a spherical truncated NFW profile, with the mass and scale radius related by $r_\text{scale}\propto M_\text{LMC}^{0.6}$ (this keeps the mass profile in the inner part of the LMC nearly independent of the total mass and satisfies the observational constraints, as illustrated by Figure~3 in \citealt{Vasiliev2021b}). We then construct equilibrium models of both galaxies, using the Eddington inversion formula for the DF in a spherical case, or the Schwarzschild orbit-superposition method for non-spherical disc+halo models (both approaches are provided by \textsc{Agama}). The MW is sampled with $10^6$ particles and the LMC -- with $0.5\times10^6$, regardless of the actual mass ratio. In disc+halo MW models, 80\% particles are allocated to the halo component. We add a few thousand tracer particles to the MW snapshot, which are sampled from a particular DF in equilibrium with the total potential, and represent the globular cluster and satellite populations, with parameters chosen to closely mimic the observed datasets. Table~\ref{tab:mock_tracers} lists the parameters of the fiducial model, which we discuss in more detail below. We considered a few additional models with the same tracer populations, LMC masses between 1 and $2\times10^{11}\,M_\odot$, and lowering the MW halo mass or adding a stellar disc; the parameters of one such model fitted to the Sagittarius stream are listed in Table~1 of \citet{Vasiliev2021b}.

\begin{table}
\caption{Parameters of the fiducial mock dataset. First two columns define the tracer populations (globular clusters and satellite galaxies), which are the same in other mock datasets; third column defines the MW (spherical halo-only model), last column defines the LMC.
The density of all components follows a tapered double-power-law profile (Equation~\ref{eq:spheroid_density}), and the DF is given by the \citet{Cuddeford1991} inversion formula with the central anisotropy coefficient $\beta_0$ and anisotropy radius $r_\text{a}$.
}  \label{tab:mock_tracers}
\begin{tabular}{lrrrr}
& GC & sat & MW halo & LMC \\
\hline
total mass [$10^{12}\,M_\odot$] & \multicolumn{2}{c}{negligible} & 1.1 & 0.15 \\
scale radius $r_\text{scale}$ [kpc] & 5 & 100 & 5 & 10.8 \\
cutoff radius $r_\text{cut}$ [kpc] & \multicolumn{2}{c}{$\infty$} & 290 & 108 \\
cutoff strength $\xi$ & \multicolumn{2}{c}{n/a} & 2 & 2 \\
inner slope $\gamma$ & 0.0 & 0.5 & 1.0 & 1.0 \\
outer slope $\beta$ & 6.0 & 6.0 & 3.0 & 3.0 \\
transition steepness $\alpha$ & 0.5 & 2.0 & 0.5 & 1.0 \\
central anisotropy $\beta_0$ & 0.0 & $-0.4$ & 0.0 & 0.0 \\
anisotropy radius $r_\text{a}$ [kpc] & 25 & 200 & \multicolumn{2}{c}{\makebox[6mm]{}$\infty$} \\
\hline
\end{tabular}
\end{table}

We then run a conventional $N$-body simulation of the encounter between the MW and the LMC, using the code \textsc{gyrfalcON} \citep{Dehnen2000}. We choose the initial position and velocity of the LMC in such a way that it arrives at the observed point%
\footnote{We use the present-day position and velocity of the LMC determined by \citet{Luri2021} and references therein: $\alpha^\text{LMC}=81.28^\circ$, $\delta^\text{LMC}=-69.78^\circ$, $\mu_\alpha^\text{LMC}=1.858\pm0.02$~\masyr, $\mu_\delta^\text{LMC}=0.385\pm0.02$~\masyr, distance $D^\text{LMC}=49.5\pm0.5$~kpc, line-of-sight velocity $v_\text{los}^\text{LMC}=262.2\pm3.4$~\kms. \label{footnote:LMCcoords}}
after 2~Gyr of evolution. The initial guess for the starting point is given by integrating the equations of motion of two galaxies (Equation~\ref{eq:lmc_mw_orbit}), and then we iteratively refine it by running a suite of 6 simulations with slightly different initial conditions, computing the Jacobian of transformation between the start and end points, and using the Gauss--Newton method to update the initial point, as described in Section~3.2 of \citet{Vasiliev2021b}. The final LMC position and velocity matches the actual observations to $\sim0.1$~kpc and $\sim0.5$~\kms. We extract the trajectories of the MW and LMC centres from the $N$-body simulation (using centre-of-mass position and velocity iteratively determined from particles within the central 10~kpc for each galaxy), fit a smooth spline curve to the former and differentiate it twice to obtain the acceleration of the MW-centered non-inertial reference frame. The positions and velocities of tracer particles at the final snapshot are used to construct the ``present-day'' mock dataset, and we also use their initial coordinates to test the method on the unperturbed MW system.

\subsection{LMC orbit reconstruction}  \label{sec:test_lmc_orbit}

\begin{figure*}
\includegraphics{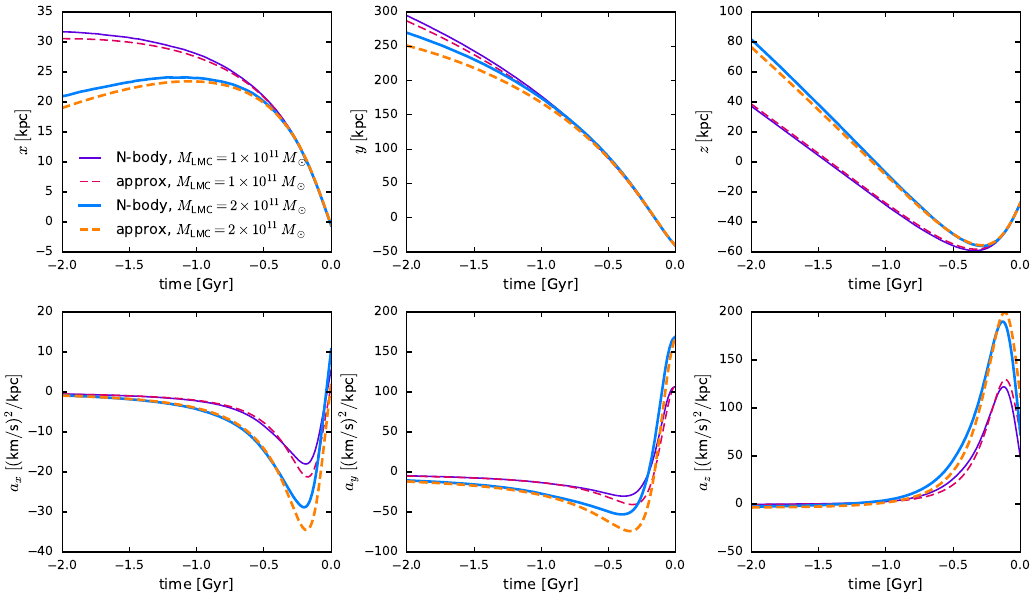}
\caption{Reconstruction of LMC's past orbit with the approximate solution of two extended bodies' equations of motion. Top row show the three cartesian components of relative distance between the MW and LMC centres, bottom row shows the components of Milky Way's acceleration. Solid lines are taken from a live $N$-body simulation, dashed lines are the orbits reconstructed via Equation~\ref{eq:lmc_mw_orbit}. Thinner and thicker lines show two cases: $M_\text{LMC}=10^{11}\,M_\odot$ and $2\times10^{11}\,M_\odot$ respectively; the MW potential is the same in both cases. Overall, the approximate solution reproduces the actual orbits reasonably well over the relevant range of LMC masses, at least over the last Gyr when the effect of the LMC flyby is most significant, although the MW acceleration is slightly overestimated by the approximation.
}  \label{fig:lmc_orbit_reconstruction}
\end{figure*}

The first question to be explored quantitatively is the accuracy of the LMC orbit reconstruction with our orbit-rewinding procedure (Equation~\ref{eq:lmc_mw_orbit}). Figure~\ref{fig:lmc_orbit_reconstruction} shows the time evolution of three components of the relative separation between LMC and MW centres (top row) and three components of the Galactocentric reference frame acceleration (bottom row), for two simulations with the same MW model but LMC masses of $10^{11}$ and $2\times10^{11}\,M_\odot$. The only adjustable parameter in this rewinding procedure is the Coulomb logarithm $\ln\Lambda$. By examining these and a few other simulations, we settle on the following expression, which provides an adequate match to $N$-body orbits in all cases: $\ln\Lambda = \ln \big[ D_\text{LMC}(t) / \epsilon_\text{LMC} \big]$. Here $D_\text{LMC}(t)$ is the instantaneous distance between the LMC and the MW centre, and $\epsilon_\text{LMC}$ is the effective ``softening radius'', for which the value $2\times r_\text{scale}$ appears to be optimal (it varies between 17 and 26 kpc for the two limiting values of LMC mass we considered). This functional form was also used by \citet{Jethwa2016} in the same context, although they found that their $N$-body trajectories were better matched by a somewhat lower $\epsilon_\text{LMC}$ in the Coulomb logarithm.

As seen from the above figure, the LMC rewinding prescription adequately recovers the time evolution of the MW--LMC separation in $N$-body simulations, but overestimates the MW acceleration around the its peak by $\sim20$\% (40\%) for the $x$ ($y$) components. This could be attributed to the neglect of the developing deformation of the MW: we implicitly assume that the entire Galaxy is pulled towards the LMC with spatially uniform acceleration, but in reality the swinging motion of the inner MW towards the LMC is partly counteracted by the gravitational force from the outer MW halo, which reacts more slowly (see \citealt{Vasiliev2021c} for an in-depth discussion). Although not perfectly reconstructed, the $x$- and $y$-components of MW acceleration change sign during the interaction, and the accumulated changes in the MW reflex velocity are smaller than in the $z$-component ($\sim20$~\kms vs. $\sim 50$~\kms, see Figure~10 in \citealt{Vasiliev2021b}); the latter is the dominant cause of the kinematic asymmetry illustrated later in Figure~\ref{fig:deltavz}, and is recovered to within a few \kms.

It is also interesting to note that the effect of changing the LMC mass is sublinear, and that the more massive LMC actually needs to start at a smaller distance in order to arrive to the same present-day point. This contrasts a simple picture used in earlier studies, in which the back-reaction of the LMC on the MW would be neglected, and LMC would move in a fixed external potential subject to additional dynamical friction force: in this case, more massive LMC would lose more energy and would have to start at a less bound orbit (i.e. further out and with a higher inward velocity), as seen, for instance, in Figure~9 of \citet{Kallivayalil2013}. Our orbit-rewinding scheme follows the mutual forces and motion of both galaxies, and is thus able to reproduce this nonlinear effect, but will likely break down for even higher LMC masses. For this reason, we put an upper limit of $2.5\times10^{11}\,M_\odot$ for $M_\text{LMC}$ in the fit, but fortunately, the best-fit models do not hit this boundary.

\subsection{Orbit rewinding in the time-dependent potential}  \label{sec:test_orbit_rewinding}

\begin{figure}
\includegraphics{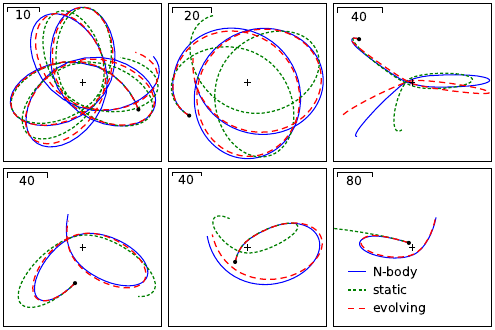}
\caption{Illustration of the accuracy of orbit rewinding for several representative orbits. Black cross is the MW centre; black dots mark the present-day positions; blue solid lines show the trajectories of particles in the original $N$-body simulation; red dashed lines are orbits integrated back in time from the present-day snapshot in the evolving MW+LMC potential; and green short-dashed lines are the orbits integrated in a static MW potential. The latter are often very different from the true ones, at least in the outer parts of the Galaxy, while the rewinding in a time-dependent potential reproduces the original trajectories $\sim 5\times$ better.
}  \label{fig:orbit_rewinding}
\end{figure}

\begin{figure}
\includegraphics{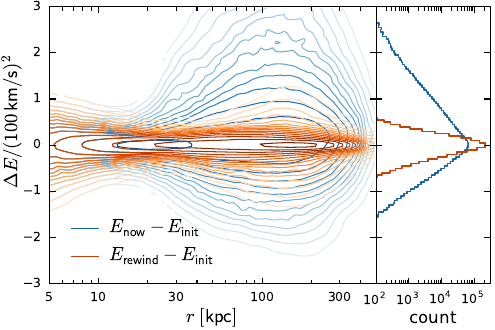}
\caption{Illustration of the accuracy of orbit rewinding. Shown are density plots of particle energy changes as a function of radius; contours are logarithmically spaced by factors of two in density. Blue contours show the changes in particle energies due to the LMC perturbation in the $N$-body simulation. Red contours show the difference between the energy of the reconstructed initial snapshot (rewinding orbits from the present-day state backward in time in the evolving analytic MW+LMC potential) and the actual initial energy of the same particles in the simulation. In the ideal case, this difference would be zero; in practice, it has some residual scatter, but much narrower than the actual energy changes. The imperfect reconstruction is likely caused by inaccuracy of the original $N$-body simulation (numerical heating in the inner few kpc, where the LMC plays no role) and the neglect of MW deformation in the orbit rewinding scheme at larger radii. 
}  \label{fig:orbit_rewinding_energy}
\end{figure}

After confirming that the LMC trajectory can be reconstructed reasonably well, we now turn to the second part of the orbit rewinding scheme -- integrating the tracer orbits back in time from their present-day coordinates in an evolving potential composed of a moving LMC, fixed analytic potential of the MW, and time-dependent but spatially uniform acceleration of the Galactocentric reference frame. Figure~\ref{fig:orbit_rewinding} shows a few typical orbits: the original particle trajectories in the $N$-body simulation are reasonably well approximated by the time-dependent orbit rewinding, but much less well by orbits in a static MW potential alone. Although some phase differences between the original and the reconstructed orbit inevitably accumulate over time, these are unimportant for the purpose of DF fitting, which uses only the integrals of motion but not phases. The relative error in position grows approximately linearly with lookback time and is at the level $\sim 0.1$ (median) for the time-dependent orbit reconstruction, while being $\sim 5$ times higher for the static potential.

Figure~\ref{fig:orbit_rewinding_energy} presents a different view on the LMC perturbation, showing the energy changes of particles in the actual $N$-body simulation at different radii (blue contours). These changes are relatively small and likely caused by numerical relaxation effects in the inner part of the system, but are much more prominent beyond $20-30$~kpc (here the horizontal axis is not the instantaneous radius of a particle, but the radius of a circular orbit with the same energy, i.e., close to a time-averaged orbit size). The energy of individual particles may increase or decrease, but on average it increases more often (note the asymmetry of the 1d histogram of $\Delta E$ in the right panel). The orbit rewinding in a time-dependent potential recovers the initial particle energies fairly well, as shown by a much tighter and symmetric distribution of the difference between reconstructed and actual initial energies (the r.m.s.\ error in energy is $\sim 10^3\,\big[\text{km\,s}^{-1}\big]^2$ for the time-dependent reconstruction and $\sim 5$ times higher in the static potential). Since the most bound orbits are little affected by the LMC, we may save effort by performing rewinding only for orbits with energies corresponding to the radii of circular orbits $\ge 10$~kpc, safely including the entire region of significant perturbations shown by blue contours in that figure.

Having demonstrated the accuracy of the orbit rewinding scheme, we now scrutinize the performance of the DF fitting approach itself.

\subsection{Measurement of the MW potential}  \label{sec:test_potential_inference}

\begin{figure*}
\includegraphics{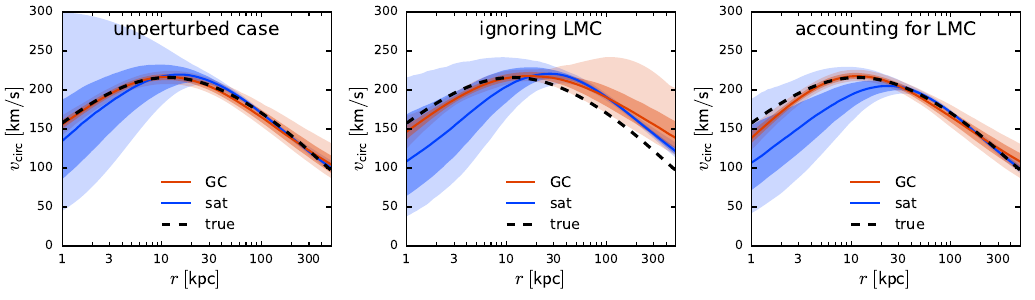}
\includegraphics{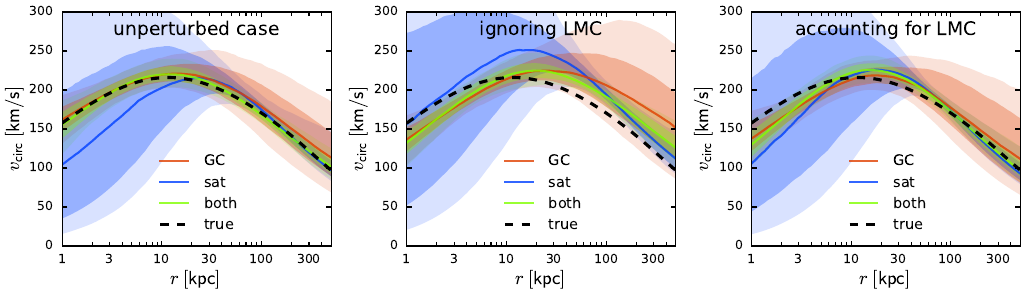}
\caption{Tests of the method on mock datasets, described in Section~\ref{sec:mock_variants}: top row shows large datasets (1000 objects in each), bottom row -- realistic sizes (150 globular clusters and 50 satellites). Each panel shows the inferred circular-velocity profiles $v_\text{circ}(r) \equiv \sqrt{G\,M(<r)/r}$ for two mock catalogues, representing globular clusters (red) and satellites (blue), and in the bottom row, the combined sample (green); the true profile is plotted by a black dashed curve. Solid lines show the medians, while darker and lighter shaded regions indicate 16/84 and 2.3/97.7 percentiles (``1$\sigma$'' and ``2$\sigma$'' confidence intervals). Naturally, the constraints are tightest in different radial ranges for the two datasets: around $5-10$~kpc for clusters and 100~kpc for dSph. \textbf{Left panel} shows the initial, unperturbed system before the arrival of the LMC, which illustrates excellent performance of the method in the ideal case. \textbf{Centre panel} shows the fit for the present-day snapshot perturbed by the LMC, but without accounting for it in the model, which leads to an overestimate of enclosed mass beyond a few tens kpc. \textbf{Right panel} demonstrates that with the orbit rewinding scheme in place, the model is able to recover the true potential.
}  \label{fig:vcirc_mock}
\end{figure*}

We ran the fitting routine for several choices of the MW potential and LMC mass, which all displayed similar performance. In this section we discuss the fiducial case of a $1.5\times10^{11}\,M_\odot$ LMC in a spherical MW potential with the total mass $1.1\times10^{12}\,M_\odot$, using either the globular cluster sample, the satellite sample, or both. For each mock dataset, we consider three scenarios: (1) initial, unperturbed MW potential; (2) present-day MW affected by the LMC, but ignoring it during the fit; (3) present-day MW with orbit rewinding compensating the LMC perturbation. 

Figure~\ref{fig:vcirc_mock}, top row, shows the results of these experiments on the full sample of tracers (1000 objects in each dataset). These ideal conditions illustrate that each of the two datasets produces tight constraints around the median radius of tracer points ($\sim 5$~kpc for globular clusters and $\sim 100$~kpc for satellites), and uncertainties increase considerably outside the range spanned by tracers, which is natural to expect if the model for the potential is flexible enough. In the unperturbed case, the mass distribution is recovered very well, and applying the vanilla DF-fitting method to a perturbed snapshot overestimates the mass at radii $r\gtrsim 30$~kpc by $\sim20-30$\%, in agreement with \citet{Erkal2020b}. Most interestingly, the addition of the orbit-rewinding step brings the inferred mass profile quite close to truth, which is very encouraging. We note that the inference is still slightly biased at $r\lesssim 10$~kpc, but at these radii the MW potential is better measured with alternative methods. 

As expected, the uncertainties grow considerably when we reduce the tracer sample to a more realistic size: 150 clusters and 50 satellites. The inferred mass profile becomes rather sensitive to the choice of the random sample (i.e., to the Poisson noise), although still remains statistically consistent between different samples due to large uncertainty intervals. The bottom row of Figure~\ref{fig:vcirc_mock} shows one of the least biased samples, illustrating the same trends as for the large sample test; in other cases, the median inferred profiles may shift up or down by a few tens percent, but the relative effect of adding the LMC perturbation and accounting for it during the fit remains the same. When we perform the fit simultaneously for both datasets, the uncertainties ``take the best of both worlds'' and become rather small in the entire radial range, also stabilizing the results against random fluctuations caused by a small sample size.

Finally, we consider the effect of adding measurement errors: 0.1 for the distance modulus (4.7\% relative distance error), 0.05~\masyr for PM, and 2~\kms for the line-of-sight velocity. These are relatively small errors, but if we ignore them in the fit, the inferred mass is biased up by 12\% at 100~kpc and by 30\% at 200~kpc. On the other hand, taking them into account restores the unbiased inference at large radii, whine increasing the uncertainty intervals only slightly (by $10-20$\%): in other words, the finite sample size (Poisson noise) is far more important than measurement errors. The green curves in the bottom panel show the fits to the combined dataset with measurement errors, which still have fairly small uncertainty intervals and recover the enclosed mass profile to within $\sim10\%$ at all radii.

The average log-likelihood of models without orbit rewinding (i.e., with biased potential inference) is lower than models with rewinding by $\Delta\ln\mathcal L\sim 5-10$ for the cluster sample and $10-15$ for the satellite sample, i.e., the non-zero LMC mass is preferred at high significance. In fact, the relative uncertainty on $M_\text{LMC}$ is at the level of 30\% when using only satellites, and twice larger when using only clusters; in the combined fit, satellites dominate. The best-fit LMC mass agrees with the true value within uncertainties.

The above results were obtained for a spherical MW model and using a \texttt{QuasiSpherical} DF of tracers. We also ran the analysis pipeline on the same mock dataset, but using \texttt{DoublePowerLaw} action-based DF (Equation~\ref{eq:DPL_DF}) in the fit (thus different from the actual DF used to construct the sample), obtaining very similar constraints for the potential and recovering the kinematic profiles of the tracer populations. Finally, additional experiments were run with non-spherical MW models taken from \citet{Vasiliev2021b} and \texttt{DoublePowerLaw} tracer DFs, again recovering the potential with good accuracy (typically within 10\% at all radii).

\section{Application to the Milky Way}  \label{sec:MW}

\subsection{Input data}  \label{sec:input_catalogues}

We use two classes of dynamical tracers: globular clusters and satellite galaxies. The former are more likely to be orbiting in the MW for many Gyr and thus are expected to be well mixed and satisfy the assumptions of equilibrium dynamical models (in absence of the LMC), but are mostly concentrated in the inner $10-20$~kpc of the Galaxy, where the LMC does not cause a significant perturbation. Satellite galaxies, on the other hand, are located much further out, at a typical distance of $\sim100$~kpc, but present an additional complication as some of them might have been accreted into the MW system only recently and thus not yet fully virialized. This certainly applies to objects at distances comparable to or exceeding the virial radius ($250-300$~kpc), such as Eridanus~II, Leo~T or Phoenix, which are still on the way towards the MW, but also possibly to galaxies that have already passed their pericentre, but still have high energy, such as Leo~I. The latter object is especially troublesome, with its distance of $\sim250$~kpc and heliocentric reflex-corrected (``Galactic standard of rest'') line-of-sight velocity of 167~\kms making it difficult to reconcile with being a long-term bound satellite, unless the MW mass exceeds $\sim1.5\times10^{12}\,M_\odot$ \citep[e.g.,][]{Kulessa1992, BoylanKolchin2013}.

For the globular cluster population, we use the most recent distance, PM and line-of-sight velocity measurements from \citet{Baumgardt2021}, \citet{Vasiliev2021a} and \citet{Baumgardt2019}, respectively. The entire catalogue contains 170 objects, but 9 of them (all at distances below 50~kpc) lack line-of-sight velocity measurements. Although datapoints with missing dimensions can still be used in model fitting, they add very little to model constraints, so we exclude them for simplicity. Furthermore, we exclude the 7 clusters associated with the Sagittarius dSph and its stream: NGC~6715 (M~54), Arp~2, Terzan~7, Terzan~8, Pal~12, Whiting~1, and NGC~2419. 

For the satellite population, we primarily use the most complete compilation of PM measurements and other properties by \citet[tables B1 and B2]{Battaglia2021}, but also rerun some of our fits with alternative PM catalogues of \citet{McConnachie2020} and \citet{Li2021}, which are also based on \Gaia~EDR3: the results were robust w.r.t.\ the choice of the PM dataset. These catalogues are not limited to the MW satellites, but contain other galaxies from the Local Group; we limit the sample to galaxies within 300~kpc, and likewise exclude objects without line-of-sight velocity measurements (although we examine their possible orbits in the potential models determined in our fit). We also remove the Small Magellanic Cloud (SMC) and other galaxies possibly associated with the LMC, as discussed by \citet{Battaglia2021}: Carina~II, Carina~III, Horologium~I, Horologium~II, Hydrus~I, Phoenix~II, Reticulum~II; again, we revisit their orbits after running the fits. Finally, we exclude a few objects that have unreliable PM measurements or are possibly unbound: Columba~I, Pegasus~III, Pisces~II, Reticulum~III. This leaves us with a sample of 36 galaxies: Antlia~II, Aquarius~II, Bo\"otes~I, Bo\"otes~II, Bo\"otes~III, Canes Venatici~I, Canes Venatici~II, Carina, Coma Berenices, Crater~II, Draco, Draco~II, Fornax, Grus~I, Grus~II, Hercules, Hydra~II, Leo~I, Leo~II, Leo~IV, Leo~V, Sagittarius, Sagittarius~II, Sculptor, Segue~1, Segue~2, Sextans, Triangulum~II, Tucana~II, Tucana~III, Tucana~IV, Tucana~V, Ursa Major~I, Ursa Major~II, Ursa Minor, Willman~1.

We propagate the measurement errors into Galactocentric positions and velocities, drawing 100 (for globular clusters) or 1000 (for satellites) Gaussian random samples from the quoted uncertainties, imposing a lower limit of 0.02~\masyr for the PM error, and 0.05 mag (for clusters) or 0.1 mag (for satellites) on the distance modulus. The results are fairly insensitive to the number of samples, since the observational uncertainties are typically quite small, with a few exceptions discussed later. However, for objects lacking line-of-sight velocities, we would have to use a larger number of samples covering a wide range of possible values, or design a more sophisticated importance sampling scheme, as in \citet{Read2021}, which is the main reason for excluding them. We also add uncertainties on the Solar position and velocity: $x_\odot=-8.12\pm0.1$~kpc, $v_\odot=\{12.9\pm1,\, 245.6\pm3,\, 7.8\pm1\}$~\kms \citep{Drimmel2018}.

\subsection{Model specifications}  \label{sec:MW_model}

We first consider the two datasets independently, running separate fits for globular clusters and satellites. In each case, we use the \texttt{DoublePowerLaw} DF family for the tracers (Equation~\ref{eq:DPL_DF}), which has 9 free parameters; simpler \texttt{QuasiSpherical} DF models are less flexible and produce noticeably worse fits (with a difference in log-likelihood $\Delta \ln \mathcal L \gtrsim 5$ for dSph alone), justifying the extra 3 parameters responsible for a non-spherical tracer density distribution and rotation. The MW halo is represented by a spherical \citet{Zhao1996} model (\texttt{Spheroid} density profile, Equation~\ref{eq:spheroid_density}) with 5 free parameters: inner and outer slopes $\gamma$, $\beta$, transition steepness $\alpha$, scale radius $r_0$ and corresponding density $\rho_0$. The outer slope $\beta$ is allowed to be shallower than 3 (the value for the NFW profile), which formally corresponds to an infinite mass (we do not impose exponential truncation in these models); as long as $\beta>2$, the potential still tends to zero at infinity, and the entire procedure remains valid.
In contrast to many other studies \citep[e.g.,][]{Eadie2019,Deason2021}, we do not artificially narrow down the prior range on any of the model parameters based on some cosmological simulations, but consider widest possible range of models and let the data speak for itself. However, to explore the sensitivity of results to the chosen parametrization of the halo density, we additionally ran another series of models with the exponential \citet{Einasto1965} profile, which has 3 free parameters: cutoff radius $r_\text{cut}$, steepness $\xi$ and normalization $\rho_0$.
We experimented with non-spherical, constant-axis-ratio models, but found that the axis ratio $q=z/x$, although not well constrained, prefers values closer to unity. Since the St\"ackel fudge is designed only for oblate or spherical potentials, we could not check if prolate models would provide a better fit%
\footnote{Note that \citet{Posti2019} also used St\"ackel fudge in conjunction with action-based DF for the globular cluster population, but did not impose the restriction $q\le 1$ on the axis ratio; thus their inferred value of $q\simeq1.3$ violated the applicability of the method and cannot be trusted.}.

We fix the parameters of the baryonic components to the values taken from the \citet{McMillan2017} best-fit potential. The bulge is a flattened truncated power-law model with $\gamma=\beta=1.8$, $\alpha=1$, $r_\text{cut}=2.1$~kpc, axis ratio $q=0.5$ and total mass $M=0.9\times10^{10}\,M_\odot$. We use a single exponential disc component in place of two stellar and two gas discs to simplify and speed up computations; it has a total mass $5.6\times10^{10}\,M_\odot$, scale radius 3~kpc and scale height 0.3~kpc. In principle, one could allow these parameters to vary, but doing so introduces degeneracy in the total potential, which should be balanced by additional observational constraints, e.g., the dependence of vertical force on $z$ inferred from stellar kinematics in the Solar neighbourhood. Since our primary goal is to explore the potential at large scales, and because the axisymmetric assumption of the St\"ackel fudge is violated in the inner Galaxy anyway, we ignore these complications, and instead pin the circular-velocity curve at the Solar radius to $235\pm10$~\kms \citep{McMillan2017}, adding a corresponding term to the likelihood function (although the results change very little if we remove this constraint). This model setup is nearly identical to the one used in \citet{Vasiliev2019b}. We also explored the effect of adding measurements of the circular velocity from \citet{Eilers2019}: these provide very tight constraints on the circular velocity profile at radii $5\le R \le 25$~kpc, but otherwise little affect the inferred mass distribution at large radii.

The LMC mass is another free parameter with a log-flat prior and an upper limit of $3\times10^{11}\,M_\odot$, and we also consider models without LMC separately, since these run a few times faster in absence of the orbit rewinding step. The uncertainty on the present-day LMC position and velocity (see footnote \ref{footnote:LMCcoords}) is propagated into the fit, adding four more parameters ($D^\text{LMC}$, $v_\text{los}^\text{LMC}$, $\mu_\alpha^\text{LMC}$, $\mu_\delta^\text{LMC}$) with Gaussian priors centered on the measured values.
We then perform a joint fit of both population (each one described by its own DF, but with a common potential), which has 28 or 26 parameters in total (5 or 3 for the Zhao or Einasto halo potential, 1 for the LMC mass, 4 for its position/velocity, and 9 for each of the two tracer DFs).
The large number of free parameters in the model may seem excessive, and indeed not all of them are well constrained (e.g., $\alpha, \beta$ and $\gamma$ span almost the entire allowed ranges $0.2-4$, $2.1-6$ and $0-2$, respectively). However, these parameters do not have an intuitively clear physical interpretation by themselves; what matters is the overall mass distribution (for the potential) and the density and velocity dispersion profiles (for the tracer DF) that they produce. We illustrate the relation between fitted potential parameters and the enclosed mass profiles in the \hyperref[sec:covplots]{supplementary material}.

Globular clusters are not expected to provide strong constraints on the Galactic potential in the outer part (beyond $\sim100$~kpc), while satellites likewise cannot constrain the inner Galaxy well. On the other hand, the combined fit, performed here for the first time, is able to put relatively tight limits on the mass distribution in the wide range of Galactocentric distances, $10-200$~kpc. We quantify it by the enclosed mass $M(<r)$ within spherical radii $r=50$, 100 and 200~kpc, or equivalently by the circular-velocity curve in the equatorial plane: $v_\text{circ}(R)\equiv \sqrt{R\,\partial\Phi/\partial R}$ (note that in a non-spherical potential, $v_\text{circ}(r)\ne \sqrt{G\,M(<r)/r}$, but the difference between them is fairly small at large radii). To facilitate comparison with other studies, we also compute the virial mass $M_\text{vir}$ and virial radius $r_\text{vir}$; however, since we do not impose a fixed functional form such as the NFW profile, the extrapolated virial mass has a much larger uncertainty than the enclosed mass within smaller radii constrained by the tracer population kinematics.

\section{Results}  \label{sec:results}

\subsection{MW potential and LMC mass}  \label{sec:mw_potential}

\begin{figure*}
\includegraphics{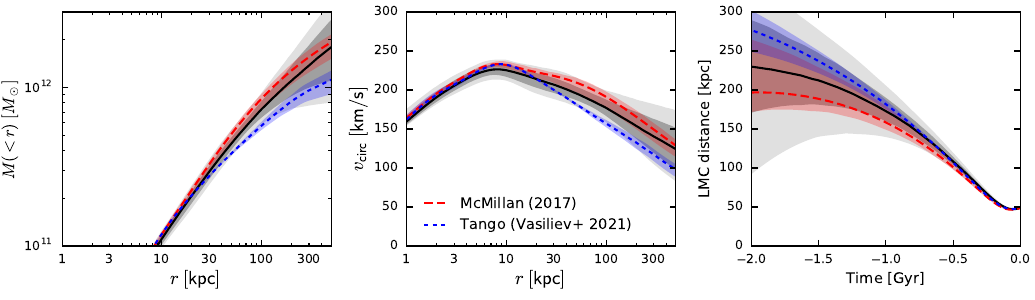}
\caption{Results of DF fits with two tracer populations (clusters and satellites), taking into account the LMC perturbation. \textbf{Left panel} shows the enclosed mass profile, \textbf{centre panel} -- the corresponding circular velocity profile, and \textbf{right panel} -- the past LMC trajectory (distance from MW centre). Median values are plotted by black solid lines, dark and light-shaded regions show the 16/84 and 2.3/97.7 percentiles. For comparison, red dashed lines and red-shaded region show the best-fit potential and an ensemble of plausible potentials from \citet{McMillan2017}, the best-fit potential being very close to the median potential in our series of models without the LMC. Likewise, blue short-dashed lines and blue-shaded region show the median and the ensemble of potentials from the Tango simulation \citep{Vasiliev2021b}, which was fit to the Sagittarius stream in the presence of the LMC.
}  \label{fig:mwpot}
\end{figure*}

\begin{table}
\caption{Constraints on the MW mass profile at different radii from various combinations of tracers (globular clusters or satellites), with or without accounting for the LMC perturbation. We list the enclosed mass at 50, 100 and 200 kpc, as well as the virial mass (for the overdensity $\Delta=102$ w.r.t.\ the cosmological matter density). The last two rows are our preferred values from both datasets with orbit rewinding, using either the baseline Zhao halo profile or the alternative Einasto profile. Distances are given in kpc and masses -- in $10^{12}\,M_\odot$, quoted as median and 16/84 percentiles.
}  \label{tab:MWpotential}
\begin{tabular}{lp{11mm}p{11mm}p{11mm}p{11mm}}
& \makebox{$M(<50)$} & \makebox{$M(<100)$} & \makebox{$M(<200)$} & $M_\text{vir}$ \\
\hline
GC, no LMC             & $0.51^{+0.12}_{-0.09}$ & $0.79^{+0.33}_{-0.20}$ & $1.08^{+0.69}_{-0.39}$ & $1.2^{+1.0}_{-0.5}$ \\[2mm]
sat, no LMC            & $0.50^{+0.07}_{-0.07}$ & $0.85^{+0.12}_{-0.10}$ & $1.31^{+0.29}_{-0.20}$ & $1.6^{+0.6}_{-0.4}$ \\[2mm]
both, no LMC           & $0.54^{+0.07}_{-0.06}$ & $0.85^{+0.11}_{-0.09}$ & $1.17^{+0.27}_{-0.23}$ & $1.3^{+0.6}_{-0.3}$ \\[2mm]
{\bf both w/LMC}       & $0.45^{+0.04}_{-0.04}$ & $0.73^{+0.09}_{-0.08}$ & $1.10^{+0.27}_{-0.22}$ & $1.3^{+0.6}_{-0.4}$ \\[2mm]
{\bf --"--, Einasto}   & $0.46^{+0.05}_{-0.04}$ & $0.74^{+0.09}_{-0.07}$ & $1.00^{+0.23}_{-0.17}$ & $1.1^{+0.4}_{-0.3}$ \\
\hline
\end{tabular}
\end{table}

We first perform separate fits for the globular cluster and satellite datasets, either neglecting the LMC perturbation or performing the orbit rewinding to compensate for it.
The MW potential inferred from the cluster population without taking LMC into account is very similar to the results of \citet{Vasiliev2019b}, which used essentially the same method but with lower-precision data from \Gaia DR2, confirming that the accuracy of observations is no longer a limiting factor. This potential is very close to the best-fit model from \citet{McMillan2017} out to a distance 100~kpc, beyond which the constraints are very weak due to absence of tracers.
The enclosed mass at 100~kpc is $0.79^{+0.33}_{-0.20}\times10^{12}\,M_\odot$ (compared to $0.85^{+0.33}_{-0.20}\times10^{12}\,M_\odot$ in the previous analysis). The fit to the satellite population provides tighter constraints at large distances, despite the smaller overall number of tracers (though we use a prior on the circular velocity at 8~kpc to pin the potential in the inner MW, which have no satellite tracers); the enclosed mass at 100~kpc is found to be $0.86^{+0.11}_{-0.09}\times10^{12}\,M_\odot$, and the median $v_\text{circ}(r)$ has a very similar profile for both datasets. Given the agreement between the two datasets, it makes sense to fit them simultaneously; unsurprisingly, the combined fit produces somewhat tighter constraints at all radii.

If we now turn on the orbit rewinding, the inferred MW mass decreases by $\sim 15\%$ at $r\gtrsim 100$~kpc, as expected from the test runs. Again the profiles are reasonably consistent between the two populations, and the combined fit to both yields the enclosed mass $0.73^{+0.09}_{-0.07}\times10^{12}\,M_\odot$ within 100~kpc; the values at other radii and for other combinations of parameters are listed in Table~\ref{tab:MWpotential}. The enclosed mass and circular velocity profiles for the fiducial series of models with LMC are shown in the left and centre panels of Figure~\ref{fig:mwpot}.

As mentioned earlier, the flexibility of the Zhao density profile means that not all of its parameters can be usefully constrained by observations, and consequently the extrapolation of the halo density profile to larger radii (and thus the inferred virial mass) depend on the allowed range of shape parameters $\alpha$ and $\beta$. Appendix~\ref{sec:covplots} shows the covariance plots between the potential parameters used in the fit and the enclosed masses at different radii (50, 100, 200 kpc and $M_\text{vir}$). We find that the outer slope $\beta$ most significantly affects the virial mass, and to a slightly lesser extent $M(<200)$, with $M_\text{vir}$ at the lower end of the allowed range of $\beta$ (2.1) being 0.1 dex (i.e., 25\%) higher than at $\beta=3$. Nevertheless, the mean likelihoods of models restricted to the range $\beta\ge 3$ is the same as for the entire range of $\beta$; thus we cannot say that there is any real preference for lower $\beta$ warranted by the data. We also examined an alternative family of Einasto profiles and found that they generally have more steeply falling density at large radii, and consequently lower virial masses (but still compatible with Zhao models within uncertainties). We conclude that in absence of more selective priors on the halo density profile, we cannot constrain the total MW mass to better than a factor of two; however, the enclosed mass at $r\lesssim 200$~kpc is better constrained for any choice of halo model, and exhibits a systematic difference of $\sim 15\%$ between the models with and without the LMC.

The median circular velocity curve in the fiducial Zhao series of models lies roughly halfway between the commonly used ``best-fit model'' for the MW potential by \citet{McMillan2017} and the potential from the ``Tango simulation'' \citep{Vasiliev2021b}, which was fit to the properties of the Sagittarius stream in the presence of the LMC perturbation. Both reference models have considerable uncertainties, which are often overlooked in the case of McMillan's potential. The difference between these two is more pronounced at large radii, and results in a qualitative difference in the inferred past orbit of the LMC (right panel of Figure~\ref{fig:mwpot}): while in the Tango simulations the LMC was nearly always on the first approach to the MW, in the heavier McMillan's potential its past trajectory would have an apocentre distance $\lesssim 200$~kpc and a period $\sim 3$~Gyr. The LMC orbits in our series of fits typically have a longer period (4--6~Gyr) and larger apocentre distance, but $\sim 90\%$ of them are still bound to the MW (i.e., had at least one additional pericentre passage in the last 7~Gyr).
However, there is a strong evidence that the LMC is on its first encounter with the MW \citep[e.g.,][and references therein]{Kallivayalil2013}. This argument alone implies a relatively light MW, on the lower end of our inferred mass range, though we caution that these models do not take into account that the MW mass was likely lower in the past, resulting in a less bound LMC trajectory for the same present mass (see Figure~11 in \citealt{Kallivayalil2013}). The gradual increase of the MW mass due to cosmological accretion of matter over many Gyr would not invalidate the DF fitting approach, since the orbital actions are conserved under slow variation of potential. Thus it is possible that the current MW mass lies in the range preferred by our fits, but the LMC is still on its first approach.

Our inferred MW mass profile agrees within uncertainties with various recent estimates (see \citealt{Wang2020} for a compilation of results). 
Before \Gaia DR2, very few satellites had PM measurements, and consequently the mass estimates varied widely between studies or even within a single study, but using different tracer subsets. For instance, \citet{Watkins2010}, using a power-law mass estimator, found $M(<300) = (0.9\pm 0.3) \times 10^{12}\,M_\odot$ using only line-of-sight velocities and assuming isotropy, or $(0.7-3.4) \times 10^{12}\,M_\odot$ under more general assumptions about anisotropy, while  the sample of 8 satellites (including LMC and SMC) with measured PM indicated a strong tangential anisotropy and hence implied $M\gtrsim 2 \times 10^{12}\,M_\odot$. The more recent PM measurements instead suggest only a mild tangential anisotropy for satellites, as discussed in the next section. With \Gaia DR2 and \textit{HST} measurements of satellite PM, and using the same power-law mass estimator, \citet{Fritz2020} measured $M_\text{vir}=(1.5\pm0.4) \times 10^{12}\,M_\odot$, or $M(<100)=(0.8\pm0.2) \times 10^{12}\,M_\odot$. The choice of method may significantly affect the result: for instance, \citet{Eadie2019}, using the same catalogue of globular clusters as \citet{Vasiliev2019b}, but employing a variant of power-law mass estimator, determined $M(<100)=0.53^{+0.21}_{-0.12}\times10^{12}\,M_\odot$, almost 40\% lower than the latter study. Recently \citet{Slizewski2021} applied the method of \citet{Eadie2019} to the satellite tracers (still using PM from \Gaia DR2) and found $M(<100)=(0.88\pm0.1)\times10^{12}\,M_\odot$. On the other hand, \citet{Wang2021} used a nearly identical method to \citet{Vasiliev2019b} with the updated \Gaia EDR3-based PM measurements of globular clusters, and obtained a rather low virial mass $0.83^{+0.36}_{-0.21}\times10^{12}\,M_\odot$ (quoted for the overdensity of 200; to compare with our definition, these values should be increased by $\sim 16\%$) when using our PM measurements with the Zhao density profile, and even considerably smaller when using the Einasto profile. They additionally imposed tight constraints on the circular velocity at $5<r<25$~kpc taken from \citet{Eilers2019}, which are based on kinematics of stars in the Galactic disc. We repeated our analysis while imposing the same prior on the inner circular-velocity curve and got very similar results to \citet{Wang2021} when using only globular clusters; however, the fits to the combined dataset of clusters and satellites instead yielded higher virial masses, more consistent with our default setup (which does not use additional priors from stellar kinematics). As we do not expect that clusters alone can provide useful constraints for the enclosed mass at $r\gtrsim 100$~kpc, we tend to trust the combined fits more, but note that even these do not tightly constrain the virial mass without additional priors on the behaviour of halo density at large radii.

Since the assumption of virial equilibrium may be questionable for the satellite population, numerous studies instead estimated the MW mass by comparing the satellite distribution with cosmological simulations: in particular, \citet{Cautun2020} find $M(<100) = 0.65^{+0.08}_{-0.06}\times10^{12}\,M_\odot$, and \citet{Li2020a} find $M(<100) = (0.72\pm0.1) \times10^{12}\,M_\odot$; these values are well compatible with our measurements. 
The MW mass profile may also be constrained by other dynamical tracers such as distant halo stars or stellar streams. A recent study of \citet{Shen2021}, using $\sim 170$ stars from the H3 survey and the \citet{Eadie2019} modelling method, measured $M(<100) = (0.69\pm 0.05)\times10^{12}\,M_\odot$, but noted that the simple power-law model cannot simultaneously fit well the density and velocity distribution of tracers. With a compilation of $\sim500$ MW halo stars from several surveys and again using a power-law mass estimator, \citet{Deason2021} found $M(<100) = 0.61 \pm 0.03\,\text{(stat.)} \pm 0.12\,\text{(sys.)} \times 10^{12}\,M_\odot$, somewhat lower than we measure. Interestingly, although they approximately accounted for the LMC influence by adding a velocity offset to the measured values, this had very little effect on the results; their systematic uncertainty rather reflects the scatter in assembly histories of galaxies in cosmological simulations.  Finally, we note that the ``Tango'' MW+LMC model fitted to the Sagittarius stream \citep{Vasiliev2021b} also has a lower mass: $M(<100) = (0.56 \pm 0.04) \times 10^{12}\,M_\odot$. Nevertheless, these values are still within the 95\% confidence interval of models in this study (in particular, if we run our fits with a fixed Tango potential, varying only the parameters of tracer DFs, the log-likelihood is decreased by $\Delta\ln\mathcal L\simeq 3$).

\begin{figure}
\includegraphics{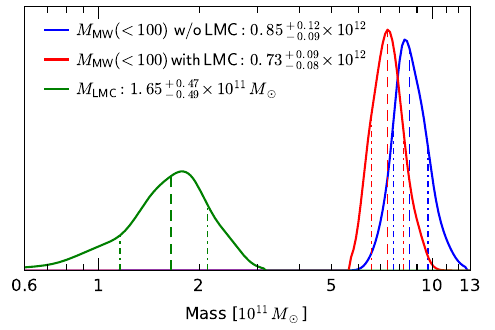}
\caption{
Posterior distribution of the enclosed MW mass at 100~kpc in models without LMC (blue) and with LMC rewinding (red), and the LMC mass in the latter series of models (green).
}  \label{fig:mass}
\end{figure}

\begin{figure}
\includegraphics{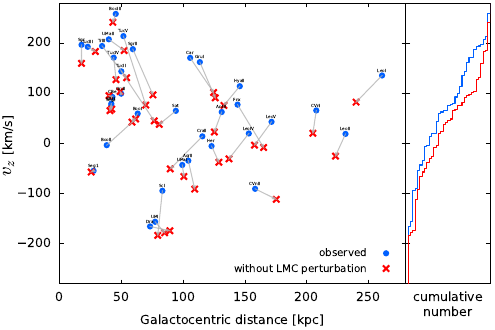}
\caption{
Illustration of the significant kinematic asymmetry in the satellite distribution induced by the LMC. Blue dots show the $z$-component of velocity plotted against Galactocentric distance for the 36 satellites selected for the fit; three quarters of them have $v_z>0$, with a mean value of 60~\kms for the entire population (see also Figure~6 in \citealt{Erkal2020b}). Red crosses show the same quantities that these object would have in absence of the LMC perturbation, obtained by integrating the present-day positions and velocities backward in time for 2~Gyr in the time-dependent MW+LMC potential and then integrating them forward in the static MW-only potential. Almost all galaxies would have had lower $v_z$ in this case, bringing the overall distribution closer to symmetry (compare the cumulative distributions in the right panel) with a mean $v_z$ of only 20~\kms.
}  \label{fig:deltavz}
\end{figure}

Although the inferred MW mass profile and the virial mass in models with or without LMC is compatible within uncertainties, the former produce materially better fits. The difference in log-likelihood is $\Delta\ln\mathcal L\simeq 12$, split roughly equally between clusters and satellites: in other words, a nonzero LMC mass is preferred with high significance. Figure~\ref{fig:mass} shows the posterior distribution of LMC mass values, which lies between $10^{11}$ and $2\times10^{11}\,M_\odot$ -- same range as inferred from various independent pieces of evidence in recent works, e.g., perturbations to stellar streams \citep{Erkal2019,Vasiliev2021b,Shipp2021}, the census of its satellites \citep{Erkal2020a,Battaglia2021}, the requirement to gravitationally bind the SMC \citep{Kallivayalil2013}, and finally, the density and kinematic asymmetries in the MW stellar halo \citep{Erkal2021,Petersen2021,Conroy2021}. The physical reason why our fits are able to constrain the LMC mass is illustrated by Figure~\ref{fig:deltavz}, which shows the measured $z$-component of velocities of satellite galaxies (for simplicity, without uncertainties). As clear from this plot and from a similar Figure~6 in \citet{Erkal2020b}, there is a significant kinematic asymmetry: the mean velocity is $\sim 60$~\kms, with three quarters of objects having positive $v_z$. If we rewind their orbits back in time in the presence of the LMC perturbation, and then integrate forward up to present time in a static MW potential, almost all of them would have lower $v_z$, decreasing the mean value threefold. Since the DF fitting (or any other method based on the equilibrium assumption) implicitly assumes a uniform distribution in orbital phases, it naturally returns a higher likelihood for a more symmetric distribution of $v_z$, and since the dispersion of values also decreases, so does the inferred MW mass.

\subsection{Properties of tracer populations}  \label{sec:satellite_orbits}

\begin{figure*}
\includegraphics{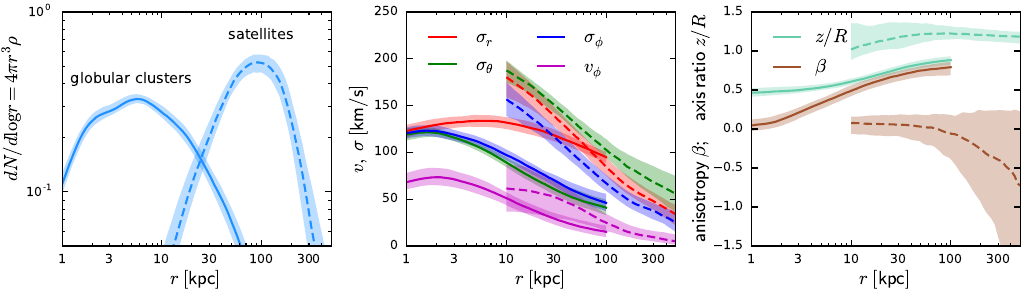}
\caption{Properties of the two tracer populations: globular clusters (solid lines) and satellites (dashed lines). \textbf{Left panel} shows the sphericalized 3d density multiplied by $r^3$, i.e., the number of objects per logarithmic interval of radii (normalized to unity for each population; the actual number of objects used in the fit is 154 clusters and 36 satellites). \textbf{Centre panel} shows the three components of velocity dispersion in spherical coordinates, and the mean rotation velocity. \textbf{Right panel} shows the velocity anisotropy coefficient $\beta \equiv 1 - (\sigma_\theta^2 + \sigma_\phi^2) / (2 \sigma_r^2)$ and the axis ratio $z/R$ of the axisymmetric density profile: clusters are flattened in the central part of the Galaxy and more spherical in the outer part, while satellites have a slightly prolate distribution overall. Lines show the median and shaded regions -- 16/84 percentiles in the MCMC ensemble of models.
}  \label{fig:tracerdf}
\end{figure*}

\begin{figure*}
\includegraphics{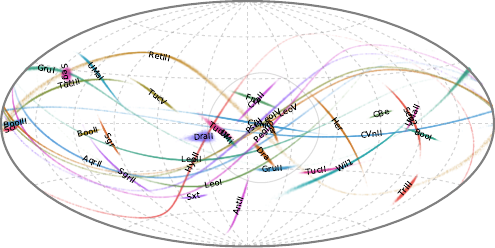}\qquad\quad
\includegraphics{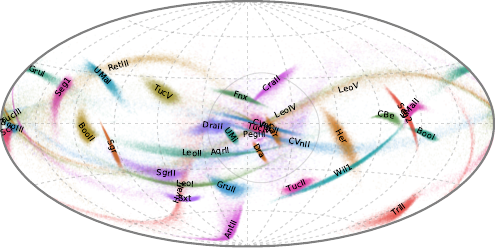}
\caption{Orbital poles of selected satellite galaxies (excluding the LMC and its likely satellites), shown on the celestial sphere in the same coordinates as Figure~3 in \citet{Fritz2018} or Figure~2 in \citet{Li2021}. Each object is shown by a cloud of points representing its posterior distribution of phase-space coordinates, with distance uncertainty usually the dominant one (hence the clouds are stretched along great circles). \textbf{Left panel} shows the actual present-day measurements, \textbf{right panel} -- the orientation of orbital poles that the objects would have if not perturbed by the LMC. The angular momentum direction of the MW disc is at the bottom of this sphere, and the largest concentration of orbital poles is around the VPOS, marked by a gray circle in the centre, which is roughly orthogonal to the MW disc.
}  \label{fig:orbital_poles}
\end{figure*}

As a by-product of our fits, we can also examine the properties of the tracer DFs manifested as the density and velocity dispersion profiles. Figure~\ref{fig:tracerdf} shows both populations on the same scale: they are clearly separated in radius by more than a factor of 10, which is beneficial for the determination of the potential, as discussed in \citet{Walker2011}. The density profiles (left panel) are well approximated by a \texttt{Spheroid} model (Equation~\ref{eq:spheroid_density}) with parameters listed in Table~\ref{tab:mock_tracers}, though both populations are non-spherical (clusters are more oblate in the inner Galaxy, and satellites are slighly prolate, as shown in the right panel). Their kinematic properties are also quite different: although both are moderately rotating in central parts ($v_\phi\sim 50-70$~\kms, see, e.g., \citealt{Frenk1980} for an early analysis of cluster kinematics), clusters are close to isotropic in the inner Galaxy and become significantly radially anisotropic further out, with $\sigma_r/\sigma_{\{\theta,\phi\}} \simeq 2$, whereas for satellites, $\sigma_r$ lies between $\sigma_\phi$ and $\sigma_\theta$, with a mild tangental anisotropy.
However, we caution that these properties are determined by the DF under the assumption of dynamical equilibrium. If we instead fit simple parametric density and velocity dispersion profiles to the actual satellite catalogue, without linking them to each other in any way, the spatial distribution of objects is significantly more extended in the $z$ direction, with axis ratio $z/R$ closer to two, but the velocity anisotropy changes from mildly tangential in the inner part to nearly isotropic or even weakly radially dominated in the outer part (see e.g. \citealt{Riley2019}). These two properties apparently cannot be reconciled in a steady-state model, hence the configuration represented by the DF is a tradeoff between these conflicting requirements, and neither the shape nor the velocity anisotropy resemble the present-day configuration particularly well.

In fact, these features are likely caused by a spatially and kinematically coherent feature known as the Vast Polar Structure (VPOS; e.g., \citealt{Kroupa2005}, \citealt{Pawlowski2020}), roughly perpendicular to the Galactic disc. We recall that we excluded the LMC and its likely satellites from the sample, but even the remaining galaxies have a significantly anisotropic distribution of orbital poles, as illustrated in Figure~\ref{fig:orbital_poles}, left panel. It has been suggested by \citet{GaravitoCamargo2021} that the gravitational effect of the LMC may cause or enhance the orbital pole clustering. However, if we ``undo'' the LMC perturbation in the same way as for $v_z$, i.e., rewinding orbits in a time-dependent MW+LMC potential and then bringing them back to present time in a static MW potential, the resulting distribution of orbital poles does not significantly change (right panel) and still remains rather non-uniform; thus the LMC perturbation cannot be the main cause of the orbital pole clustering (the same conclusion is independently reached by \citealt{Pawlowski2021}). \citet{Li2021} estimated that around a half of the entire satellite population may be part of VPOS, including the LMC itself and its satellites.

In the above figure and in subsequent ones, each object is rendered as a cloud of points representing the uncertainty in its orbit parameters, which stems both from the measurement uncertainties and from the variation in the potential in our ensemble of models (including the LMC mass). However, it is important to stress that one should not sample uniformly from the measurement error distribution, but rather from the posterior distribution (i.e., weighted by the DF value). As a simple illustration, consider an object moving on a nearly circular orbit at a certain radius, and imagine that its PM is measured with a relatively large uncertainty comparable to the value itself. Then most samples from this error distribution will produce a higher velocity and hence place the object at the pericentre of an eccentric orbit with a larger apocentre radius, or even on an unbound orbit. Thus if one assigns equal probability to all these samples, one would arrive at an incorrect conclusion that the object is likely near the pericentre of a highly eccentric orbit. If, on the other hand, one accounts for the fact that lower velocity corresponding to a more tightly bound orbit is more likely to be found among the ensemble of all orbits generated by the DF, the inference will be unbiased. This resolves the apparent ``double convolution'' paradox: assuming that we have a given intrinsic distribution of some quantity (say, velocity), and that measured values are equally likely to be smaller or larger than the true value for any object, the observed distribution will be broadened by errors. Now by sampling from a Gaussian error distribution around the \textit{measured} value, we effectively broaden the true distribution a second time, but by assigning higher weight to samples that have smaller velocities, we may recover the true width of the intrinsic distribution, and the posterior distribution of possible true velocity values for each object will be shifted towards lower-than-measured values. Of course, this requires a model for the intrinsic distribution to be constructed from the measured values, which is precisely what we do in our fits. A caveat is that if the model contains only the equilibrium DF in the given potential, all posterior orbits will necessarily be gravitationally bound to the MW. However, since we introduced a second, ``unmixed'' population of object containing both positive and mildly negative energies, we do not artificially enforce the orbits to be bound.

To construct our posterior sample of orbit parameters $\boldsymbol w_i$ for the $i$-th object, we first draw samples $\boldsymbol w_i^{(s)}, s=1..N_\text{samp}$ from the Gaussian error distribution of this object. Then we iterate over a large number of model parameters from the MCMC ensemble, and for each model $m$ of the potential and the DF $f_m(\boldsymbol w)$, evaluate the likelihood of all samples:
\begin{equation}
f_m(\boldsymbol w_i^{(s)}),\quad s=1..N_\text{samp}\,,
\end{equation}
then pick up one or more samples with probability
\begin{equation}
p_i^{(s)} \equiv \frac{f_m(\boldsymbol w_i^{(s)})}{\sum_{s'=1}^{N_\text{samp}} f_m(\boldsymbol w_i^{(s')})} \,,
\end{equation}
and finally stack together the posterior samples from all examined models (note that the denominator in the above equation is the error-convolved DF value as in Equation~\ref{eq:error_convolution_sum}). For the majority of objects, the posterior distribution averaged over many models in the MCMC ensemble is actually very close to a normal distribution centered on the measured values with quoted uncertainties, i.e., the measurement errors are effectively negligibly small. One may quantify the reduction in the posterior width by the entropy, defined as 
\begin{equation}
S_i \equiv -\!\!\sum_{s=1}^{N_\text{samp}} p_i^{(s)}\; \ln\big(p_i^{(s)}\,N_\text{samp}\big) \le 0.
\end{equation}
If the measurement errors are small, all $p_i^{(s)}$ are close to $1/N_\text{samp}$, and hence the entropy is close to zero; if a single sample dominates the sum, $S\approx -\ln N_\text{samp}$. Objects with the largest negative entropy show the largest reduction in the posterior width, and unsurprisingly, typically have relatively large PM uncertainties or no line-of-sight velocity measurements. These include a few faint globular clusters, namely, Crater, AM~1, AM~4 and Mu\~noz~1: the first two are also the most distant ones, while the latter two are very faint clusters at distances $\sim30$ and $\sim45$~kpc respectively, which have no stars on the giant branch. Among satellite galaxies with $S<-0.5$ are Columba~I, Eridanus~II, Hydra~II, Leo~IV, Leo~V, Pegasus~III, Pisces~II, Reticulum~III (most of them are rather distant), and some (but not all) objects with no line-of-sight velocities: Bo\"otes~IV, Cetus~II, Cetus~III, Indus~I, Pictor~I.

In a mixture model for the satellite sample, we additionally evaluate the posterior probability of belonging to the unmixed (infalling, splashback and unbound) population for each object (Equation~\ref{eq:posterior_outlier_probability}). It turns out that in models without the LMC, Leo~I has a high probability of being unmixed, which is not surprising, given its high velocity and large distance. However, when accounting for the LMC perturbation, its pre-LMC orbit actually has a lower energy and is likely to belong to the main sample, except if the MW mass is small enough ($M_\text{vir}\lesssim 0.9\times10^{12}\,M_\odot$). Hercules, on the other hand, is confidently bound in all models without LMC, but also has a significant probability of being unmixed in a low-mass MW+LMC scenario. Although the mixture model is much less sensitive to an occasional outlier than a standard model with only bound population, we repeated the fits excluding these two galaxies from the sample. The inferred mass within 100 kpc was $\sim 10\%$ lower in this case, but well within uncertainties of the fiducial scenario.
We also looked at the probability of belonging to the unmixed population for other objects not included in the main sample. Apart from the likely LMC satellites, which are examined in the next section, the following objects were found not to be among virialized satellites: Columba~I, Eridanus~II, Leo~T, Phoenix (which are all distant and still infalling), and possibly Cetus~III and Bo\"otes~IV (which have no line-of-sight velocity measurements).

\begin{figure*}
\includegraphics{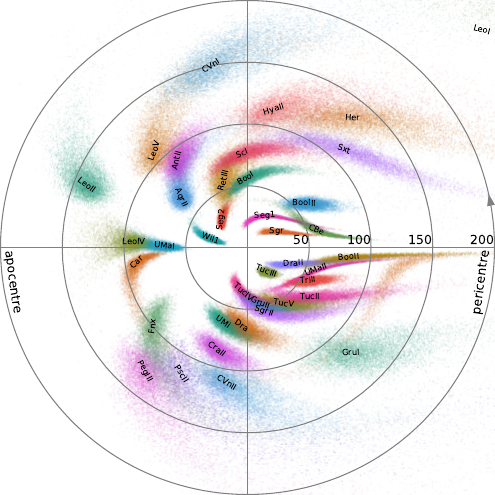}\qquad\quad
\includegraphics{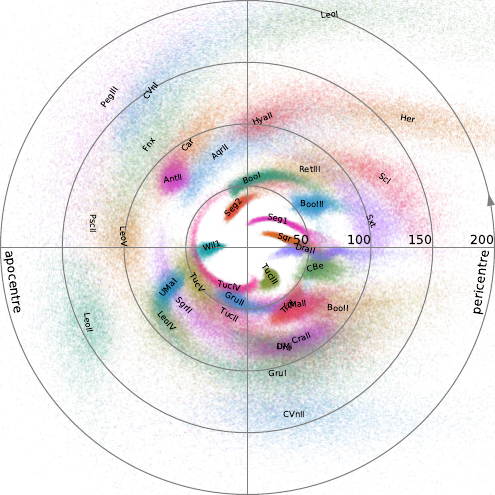}
\caption{Distribution of satellites in the space of energy and orbital phase. In the polar coordinates, radius corresponds to the radius of a circular orbit with the given energy, and polar angle -- the radial phase angle $\theta_r$; each galaxy is rendered by a cloud of points sampling from the joint posterior distribution of MW and LMC potential models and measurement uncertainties. Only a subset of all satellites is shown, excluding the galaxies associated with the LMC and a few likely non-virialized objects. \textbf{Left panel} shows the ensemble of models without the LMC, so that the location of points corresponds to the actual present-day phase-space coordinates of these objects. \textbf{Right panel} depicts the ensemble of models with the LMC: the energy and orbital phase are obtained by integrating orbits backward from the present-day coordinates in the time-dependent MW+LMC potential, and then integrating forward to present time in a static MW potential, thus the scatter of points is increased by the variation in the MW potential parameters.
}  \label{fig:orbital_phase}
\end{figure*}

With the posterior samples, we construct the diagram of orbital phases of satellite galaxies: for each choice of potential and each sample from the position/velocity distribution, we compute the radial phase angle $\theta_r$ -- a canonically conjugate variable to the radial action, which uniformly increases with time. A well-mixed population is expected to have a uniform distribution of $\theta_r$. Figure~\ref{fig:orbital_phase} shows the 36 galaxies from the main dataset used in the fit, plus a few objects with poor PM measurements that were excluded from the sample (Pegasus~III, Pisces~II, Ridiculum~III). This plot does not include the LMC itself, its likely satellites, and objects with high posterior probability of belonging to the unmixed population or without line-of-sight velocities. We observe that there is no significant bias towards close-to-pericentre orbital phases, in contrast to the analyses of \citet{Simon2018}, \citet{Fritz2018} and \citet{Li2021}, which, however, used raw measurements rather than posterior distribution, possibly leading to a bias as explained above. The inclusion of the LMC does not significantly change this distribution (we again perform a backward-then-forward rewinding to undo the LMC perturbation, but caution that the reconstruction of angle variables may be less accurate than the integrals of motion). 

However, we caution that the DF fitting method (or any other approach invoking the Jeans theorem) implicitly assumes a uniform distribution in angle space, and hence prefers potentials that make the observed population consistent with uniform random sampling of orbital phases. If the MW mass were lower, the same measured velocities of objects would be relatively larger compared to the equilibrium velocity distribution at a given radius, and hence one would infer that many of the objects are near the pericentres of their orbits. However, since there is no corresponding population of more distant and slower objects near their apocentres, this potential would be assigned a lower likelihood in the model. On the other hand, if there actually exist distant undiscovered satellites (for instance, \citealt{Koposov2008} estimate that up to $50-100$ faint satellites are still undetected), the inferred MW potential would need to be revised downwards, as illustrated by a toy experiment in the \hyperref[sec:potential_bias]{Appendix}.

\begin{figure*}
\includegraphics{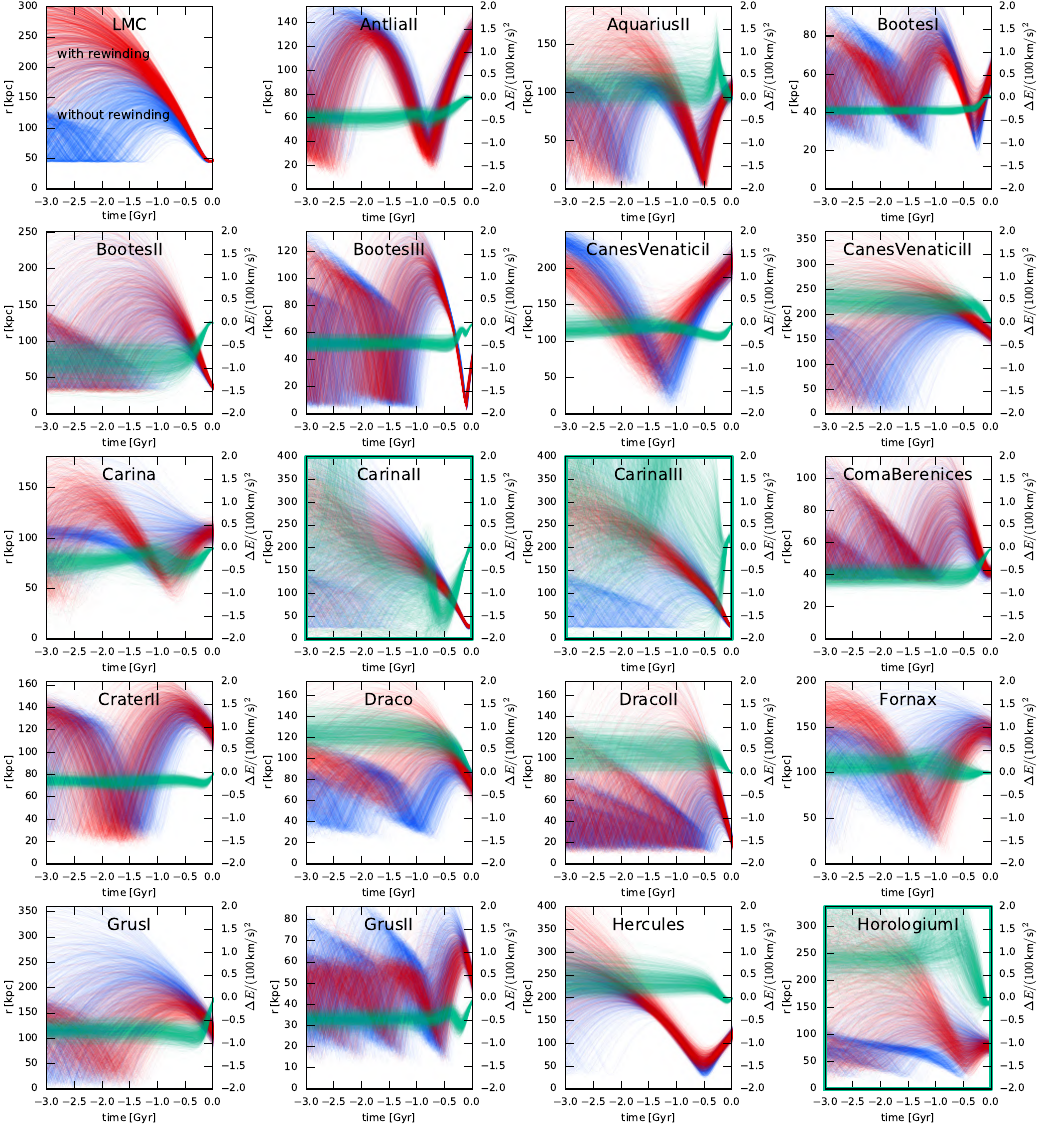}
\caption{
Orbits of satellite galaxies in the models neglecting the LMC (blue) or including it (red). For each object, we show the evolution of the Galactocentric radius for a variety of orbits integrated in different potentials with parameters drawn from the MCMC ensemble of models, and the orbital initial conditions (present-day positions and velocities) drawn from the posterior distribution of each model (i.e., are sampled from measurement uncertainties and weighted with the probability of each sample in the given DF and potential model).  The top left panel shows the trajectories of the LMC itself (neglecting observational uncertainties). In the remaining panels, cyan lines additionally show the evolution of energy for each orbit with respect to its present-day value (secondary axis) for the case of the time-dependent MW+LMC potential. When the energy does not change significantly (e.g., for Crater~II or Fornax), the orbits in the time-dependent potential are somewhat more extended since the MW is on average less massive in this series of models. But the recent LMC-induced energy kicks may be both positive or negative and can significantly alter the inferred orbits (e.g., for Carina or Grus~I), or the object might end up being bound to the LMC over its past orbit, in which case the energy in the MW potential varies wildly (e.g., Carina~III).
}  \label{fig:orbits}
\end{figure*}

\begin{figure*}
\includegraphics{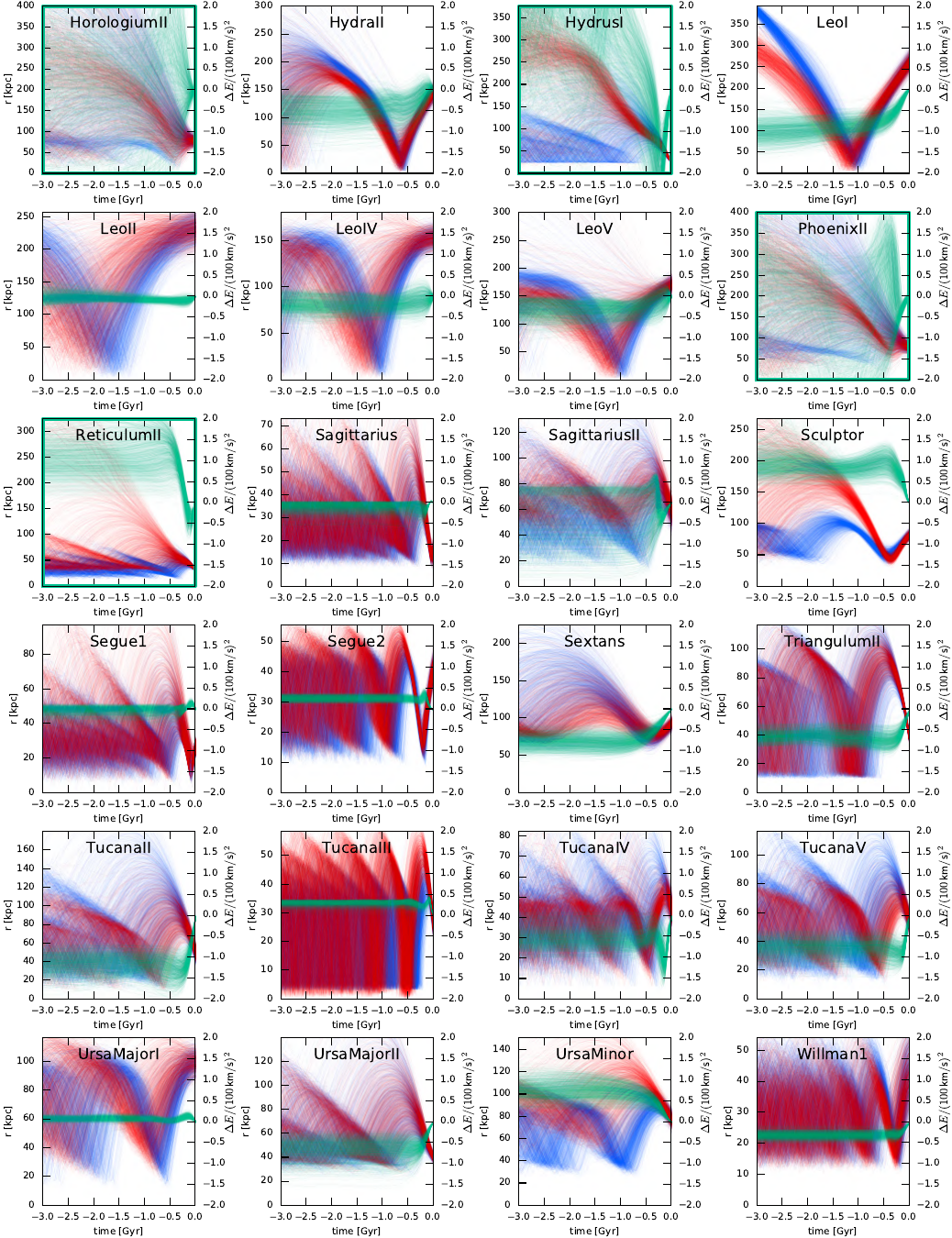}
\contcaption{}
\end{figure*}

We now examine more systematically the effect of the LMC on the orbits of clusters and satellite galaxies, using the posterior samples in two MCMC ensembles of models, with and without orbit rewinding. As expected, we do not find any significant changes in cluster orbits with or without the LMC, in agreement with \citet{Garrow2020}, nor are any cluster orbits associated with the LMC, confirming \citet{Boldrini2021}; thus we concentrate on the satellites in what follows. Figure~\ref{fig:orbits} shows the time evolution of Galactocentric distance over the last 3~Gyr%
\footnote{In this section, we reconstruct the LMC trajectory and rewind the orbits of tracers for a longer period of 3~Gyr, even though the likelihood of models is still computed based on the positions and velocities of tracers 2~Gyr ago.}
for models without LMC in blue, and with LMC in red. Green lines additionally show the evolution of orbit energy with respect to its current value in the latter series of models: if it stays around zero, the LMC has little effect on the orbit, but it can also be larger in the past (i.e., the galaxy was moved by the LMC onto a more tightly bound orbit) or vice versa. We observe that for some objects (e.g., Segue~1 or Willman~1), especially in the inner Galaxy, the red and the blue series of orbits are rather similar, but for the majority of them, the orbits look quite different. There are, of course, satellites of the LMC, which have been affected by it over the entire time interval, as evidenced by the large spread of green curves (e.g., Carina~III or Hydrus~I); we discuss them in more detail in the next section. For others, the influence of the LMC is mainly limited to the last 0.5~Gyr. We stress that this needs not be a direct gravitational scattering by the LMC itself (though this likely happened for Aquarius~II, Grus~II, Sagittarius~II, Tucana~III, Tucana~IV, Tucana~V), but rather the change in MW-centric orbit parameters as a result of the MW reflex motion caused by the LMC. As illustrated by Figure~\ref{fig:orbit_rewinding_energy}, objects beyond a few tens kpc are likely to be moved up or down in energy, but the former occurs more often. For some galaxies, this dramatically alters their inferred past orbit; a prime example is Leo~I, which had a more tightly bound orbit in the past until the arrival of the LMC, as already noted by \citet[their section 3.2]{Erkal2020b}. A similar conclusion is reached for Leo~V, Sextans and Tucana~II. On the other hand, a surprisingly large fraction of satellites have been moved to more tightly bound orbits recently, including Canes Venatici~II, Draco, Fornax, Hercules, Sculptor and Ursa~Minor. We note that in absence of any energy change (e.g., Crater~II or Ursa Major~I), orbits in the series of models with LMC would be somewhat larger than in models without LMC, on account of a lower MW mass in the former case, but the difference in orbits for the above named galaxies is far more pronounced. Among them, only Hercules is affected to such an extent that it might have been on a splashback orbit prior to the LMC-induced perturbation, but the orbits of Carina and Sculptor are also dramatically changed and had a higher eccentricity in the past. We note that the orbit of Sagittarius dSph also increased its eccentricity due to the LMC perturbation (see Figure~9 in \citealt{Vasiliev2021b}), but these changes are still minor compared to what many other satellites have experienced. It is clear that the LMC cannot be neglected when analyzing the past trajectories, tidal forces and mass loss history of most MW satellites.

\subsection{Satellites of the LMC}  \label{sec:lmc_satellites}

\begin{figure*}
\includegraphics{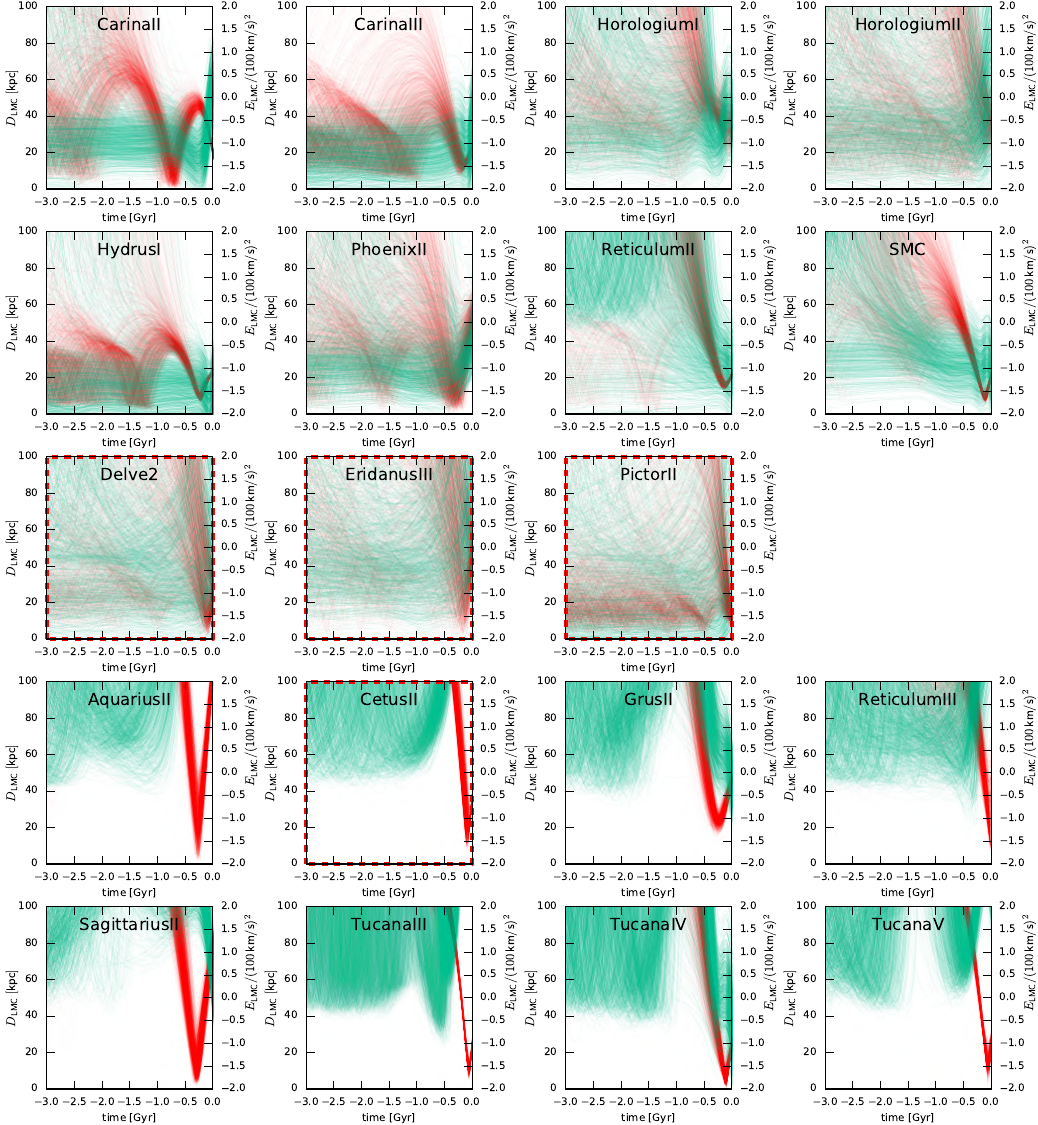}
\caption{History of interaction with the LMC for selected galaxies over the past 3 Gyr. For each object, red lines show the distances from the LMC as a function of time for an ensemble of orbits sampled from the uncertainties in the present-day position/velocity, variations in the MW potential, and the LMC mass; cyan lines show the relative energy w.r.t.\ the LMC (ignoring the MW). A substantial fraction of green lines below zero indicates that the object is likely to be an LMC satellite (though it might not be bound at present time): this is the case for the galaxies in the first two rows, though Horologium~II and Reticulum~II have only a moderate overall probability of being bound in the past. The galaxies in the third row lack line-of-sight velocity measurements, but could have been bound to the LMC for a range of possible velocities. The last two rows show galaxies that have experienced a close passage with the LMC recently, but were not bound to it earlier, and with a possible exception of Grus~II and Tucana~IV, are unlikely to be bound at present time. 
}  \label{fig:orbits_wrt_lmc}
\end{figure*}

\begin{figure*}
\includegraphics{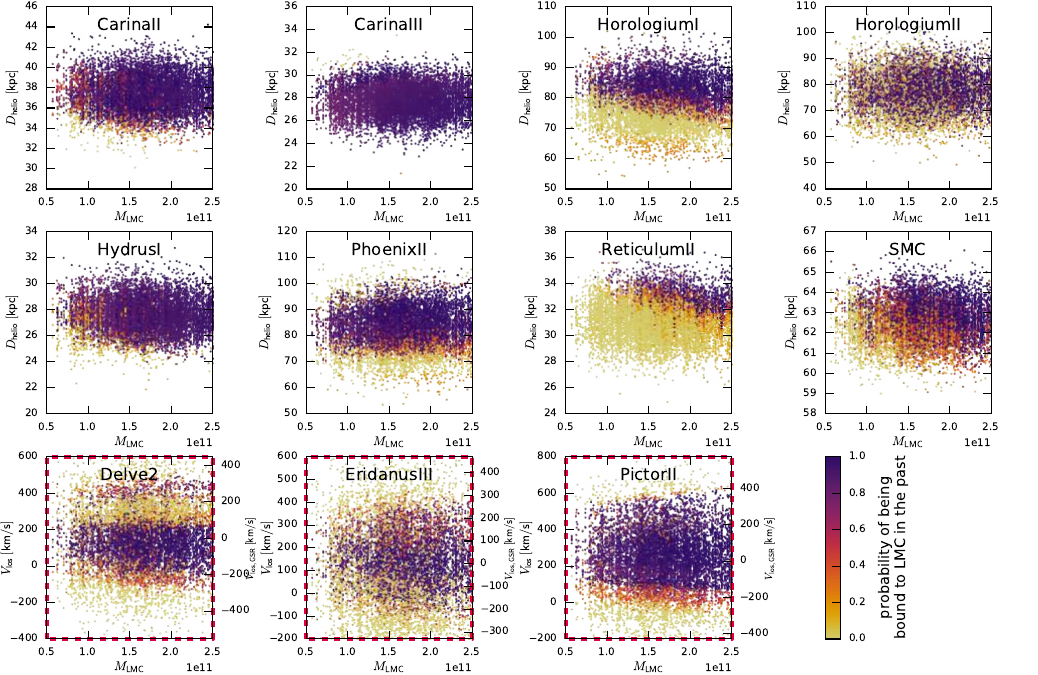}
\caption{Probability of past association with the LMC for selected galaxies. Each point represents an orbit sampled from the uncertainties in the present-day position/velocity, variations in the MW potential, and the LMC mass (\textit{without} posterior reweighting); the colour indicates the fraction of time that this orbit had a negative energy w.r.t.\ LMC in the interval between $-3$ and $-1$ Gyr in the past. Orbits are plotted against LMC mass and either the heliocentric distance (top two rows) or the presently unknown line-of-sight velocity (bottom row).
}  \label{fig:boundprob_lmc}
\end{figure*}

We now turn attention to the objects suspected to be LMC satellites and therefore excluded from the main sample during the fit (marked by green frames in Figure~\ref{fig:orbits}). Their past orbits in models with the LMC are indeed concentrated around the past trajectory of the LMC itself, and their energy in the MW-centered frame has large fluctuations at all times, not just over the last 0.5~Gyr. For a better view, Figure~\ref{fig:orbits_wrt_lmc} shows the evolution of the distance from the LMC for these and a few other galaxies, and the evolution of energy in the LMC-centered frame $E_\text{LMC}$ (i.e., using the relative velocity). Although this energy may change rather dramatically in the last 0.5~Gyr due to the complex three-body interaction between the MW, the LMC and the satellite galaxy, it often ends up being negative for a large fraction of orbits (recall that the ensemble of orbits encompasses the uncertainties in the current coordinates, MW potentials and LMC masses, but in this case we do not additionally reweigh orbits by their posterior probability in the MW-bound DF).

To quantify the probability of being an LMC satellite, we use the fraction of orbits with $E_\text{LMC}<0$ over the time interval between $-3$ and $-1$~Gyr. Figure~\ref{fig:boundprob_lmc} shows this fraction as a function of the LMC mass and a second parameter -- either the heliocentric distance or the line-of-sight velocity. We find that Carina~II, Carina~III, Horologium~I, Hydrus~I, Phoenix~II and SMC have high probability of being bound to the LMC in the past regardless of the LMC mass (though it generally increases with mass), at least in most of the range of heliocentric distances compatible with observations. Reticulum~II has a lower probability overall, unless its distance is underestimated (typically it would be bound for $D\gtrsim 35$~kpc for $M_\text{LMC} \ge 10^{11}\,M_\odot$ or $D\gtrsim33$~kpc for $M_\text{LMC} \ge 2\times10^{11}\,M_\odot$, while the observed distance is $31.6\pm1.5$~kpc). Horologium~II also has a large spread in probabilities at all LMC masses, possibly due to relatively large PM uncertainties.

In addition, we find a few LMC satellite candidates among objects with no line-of-sight velocity measurements, shown in the third row of Figures~\ref{fig:orbits_wrt_lmc} and \ref{fig:boundprob_lmc}. Pictor~II, Delve~2 and Eridanus~III are currently located within 10, 30 and 40 kpc from the LMC, respectively, but their PM are compatible with the past association with the LMC (marginally, in the case of Eridanus~III). Naturally, this is conditioned on the currently unknown velocity, namely: $v_\text{los}$ needs to be $110\pm80$~\kms for Delve~2, $200\pm70$~\kms for Eridanus~III, or $270\pm120$~\kms for Pictor~II. The association between these objects and the LMC has been previously proposed in \citet{Cerny2021}, \citet{Kallivayalil2018}, \citet{Erkal2020a} and \citet{DrlicaWagner2016}, with similar predictions for the missing velocity.

Galaxies shown in the lower two rows of Figure~\ref{fig:orbits_wrt_lmc} had a close passage with the LMC in the last few hundred Myr, but were not among its satellites in the past. Two of them (Grus~II and Tucana~IV) currently may have velocities below the escape velocity of the LMC, but might not remain bound in the future on account of strongly dynamic interactions with both the LMC and the MW. 
For the remaining galaxies, we do not find any significant probability of LMC association in the past, although some of the brightest satellites (e.g., Carina, Fornax) share the orbital plane with the LMC and might be accreted from the same filament on cosmological timescales \citep[e.g.,][]{Pardy2020}.

Our assessment of the LMC satellite population broadly agrees with other studies (\citealt{Jethwa2016}, \citealt{Kallivayalil2018}, \citealt{Erkal2020a}, \citealt{Patel2020}, \citealt{Battaglia2021}), which typically limited their analysis to one or a few choices of MW potential. For Carina~II, Carina~III, Horologium~I, Hydrus~I, SMC there is a general concensus that they have been associated with (a massive enough) LMC in the past. Horologium~II, Phoenix~II and Reticulum~II are more debatable; we find respectively moderate, high and low probability of them being former LMC satellites.

\section{Discussion and conclusions}  \label{sec:summary}

The role of the LMC as a major factor affecting the dynamics in the outer reaches of our Galaxy is gradually becoming appreciated by the community, but the tools for incorporating this factor into analysis are still scarce. Here we introduced a simple but effective approach for ``restoring order'' by first computing the past orbits of the MW and the LMC under mutual gravity, and then rewinding the orbits of stars or any other MW-bound objects in a time-dependent potential of the two galaxies, thus reconstructing the unperturbed state of the MW before the arrival of the LMC.
This opens up the possibility to apply well-developed classical methods for dynamical modelling and measurement of gravitational potential based on the equilibrium approximation. 

We then invoked one such method, simultaneously constraining the tracer DF, the MW potential, and the LMC mass, using two complementary datasets of objects with 6d phase-space coordinates: globular clusters and satellite galaxies. We demonstrated that the method is able to recover the true potential in the presence of the LMC perturbation, and that the neglect of this perturbation biases the inferred MW mass up by $\sim20$\%, in agreement with estimates by \citet{Erkal2020b}. Applying our method to the most recent measurements based on \Gaia EDR3, we measured the MW mass distribution up to the virial radius with a relative uncertainty of $\sim20\%$ at 200~kpc (see Table~\ref{tab:MWpotential}), although we do not obtain tight constraints on the virial mass. The most likely range for the LMC mass is $(1-2)\times10^{11}\,M_\odot$; models without the LMC perturbation are noticeably worse in fitting the observed dataset (the difference in log-likelihood is $\sim 12$).

We stress that the effect of the LMC cannot be simply reduced to a deflection of orbits that pass close to it. Equally, if not more important, is the reflex motion of the central region of the MW in response to the LMC's gravitational pull, displacing it spatially and kinematically w.r.t.\ the outer part. The more distant objects are not able to keep up with this recent swinging motion of the MW-centered reference frame, and thus find themselves on significantly different orbits than without the LMC perturbation: this is clearly manifested in the energy variations in the last 0.5~Gyr in a large fraction of satellite galaxies (green curves in Figure~\ref{fig:orbits}) and the corresponding differences in inferred past orbits with and without the LMC (red and blue curves in that figure). Nevertheless, the global features of the entire population of satellites, such as the orientations of angular momentum vectors (Figure~\ref{fig:orbital_poles}) or distribution of orbital phases (Figure~\ref{fig:orbital_phase}), are robust w.r.t.\ the inclusion of the LMC.

In contrast to some other studies (e.g., \citealt{Fritz2018}, \citealt{Li2021}), we do not find any evidence for a concentration of satellites near the pericentres of their orbits, nor for a strong tangential anisotropy of their velocities. We stress that this assessment is based on the limited selection of galaxies that does not include the LMC and its likely satellites, which are, of course, mostly near their pericentres, but should not count as independent samples. However, there are two further reasons for our different conclusion. One is that we use posterior-weighted uncertainties rather than raw measurements, which downweights the high-velocity tail of the observational error distribution. The other is that since we let the potential vary in the fit, the models choose the MW mass profile that prefers a more uniform distribution of orbital phases by construction. Nevertheless, we highlight a possible caveat in this line of reasoning: if the observed catalogue of satellites is incomplete and there exist yet undiscovered objects at large distances (hence more likely to be at apocentres), our inference on the MW mass would be biased up, as illustrated in the \hyperref[sec:potential_bias]{Appendix}, and in the true (lower-mass) potential, the observed distribution of satellites should indeed be concentrated near pericentres. If the selection function of the observed sample could be reliably estimated, this effect may be taken into account in our modelling scheme, as illustrated e.g.\ by \citet{Hattori2021} in a similar context.

Inevitably, our approach has a number of limitations. For simplicity, we neglect the measurement uncertainties on the LMC position and velocity, which are fairly small compared to uncertainties of many other objects. We also ignore the gravitational interaction between LMC and SMC, which might deflect the LMC and thus affect its past orbit reconstruction. More fundamentally, we assume a fixed (non-deforming) potential of both MW and LMC; note that the deformation of the tracer density profile is implicitly accounted for by the orbit rewinding procedure. Nevertheless, even this rather simplified reconstruction of the LMC orbit is able to compensate, in the first approximation, its perturbation on MW tracers.

The orbit rewinding step can be easily added to any forward-modelling approach constrained by kinematics of individual tracers, such as halo stars or tidal streams. Thus the significant gravitational perturbation from the LMC does not invalidate the classical dynamical modelling tools, but should be taken into account if we aim for a 10\% accuracy level that matches the quality of modern observational datasets.

\section*{Acknowledgements}
We thank V.Belokurov, D.Erkal, W.Evans and S.Koposov for valuable discussions, and the anonymous referee for expedient and comprehensive reports, which prompted us to extend our analysis and clarify various issues, improving the presentation. EV acknowledges support from STFC via the Consolidated grant to the Institute of Astronomy.\\
\textit{Software used}: \textsc{Numpy}, \textsc{Matplotlib}, \textsc{Emcee} (\href{http://ascl.net/1303.002}{ascl:1303.002}), \textsc{GyrfalcON} (\href{http://ascl.net/1402.031}{ascl:1402.031}), \textsc{Nemo} (\href{http://ascl.net/1010.051}{ascl:1010.051}).

\section*{Data availability}
We provide samples of the MW and LMC potential parameters from our MCMC analysis and corresponding satellite orbits, together with example \textsc{python} scripts for orbit rewinding, at the following repository: \url{https://zenodo.org/record/5541971}.

\appendix

\begin{figure*}
\includegraphics{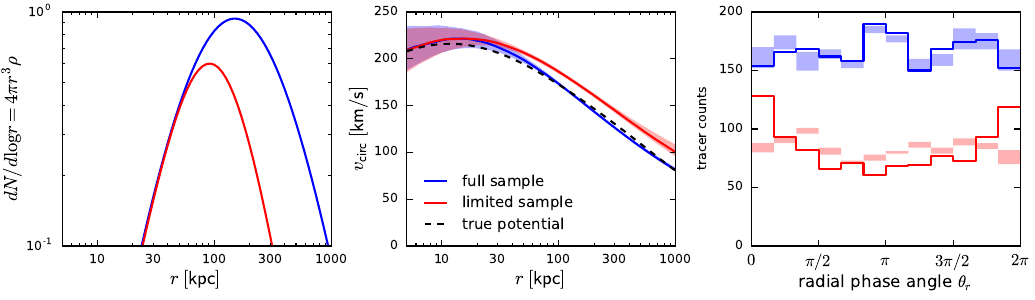}
\caption{Illustration of the potential bias caused by incomplete tracer sample (see text for details).
}  \label{fig:potential_bias_incomplete_sample}
\end{figure*}


\vspace*{-5mm}
\section{Bias in potential from incomplete tracer samples}  \label{sec:potential_bias}

Here we present a simple illustration of the bias in potential inference resulting from an incomplete tracer sample. The true distribution of tracers is sampled from an isotropic \texttt{QuasiSpherical} DF corresponding to the density profile shown in blue in the left panel of Figure~\ref{fig:potential_bias_incomplete_sample}. The limited sample is obtained by removing some of the more distant objects, so that the spatial distribution of the remaining ones follows the red curve (it matches the original sample at $r\lesssim 50$~kpc and becomes progressively more incomplete at larger radii). Naturally, this limited sample is also more likely to contain objects near pericentres of their orbits -- the distribution of radial phase angles of the limited sample in the actual potential is more peaked around $\theta_r=0,2\pi$, unlike the uniform distribution of the full sample (both shown by solid hustograms in the right panel). Although there is no explicit kinematic selection, the remaining objects have on average larger velocities than if we observed them equally spread over all orbital phases.
Centre panel shows the circular-velocity curve obtained by simultaneously fitting the DF and the potential to the full sample (blue shaded region) and to the limited sample (red shaded region). The former one recovers the actual circular velocity (black dashed line) to within 3\%, while the latter one is biased up by 7\% at 100~kpc and by 15\% at 300~kpc (the biases in the enclosed mass are twice larger). The reason is that the Jeans theorem, which underlies any equilibrium modelling approach, implies a uniform distribution in phase (red shaded regions in the right panel), and hence compensates for the pericentre bias in the limited sample by increasing the host galaxy mass, so that the higher velocities in the limited sample are representative of the average orbit velocity rather than the pericentre velocity. The DF fitted to the limited sample is also slightly more tangentially biased ($\beta\simeq -0.05$) for the same reason (the actual velocities of remaining objects near pericentre are preferentially tangential, and the assumption of uniform phase distribution propagates this to the entire orbit).

\vspace*{-4mm}
\section{Covariance plots for potential parameters}  \label{sec:covplots}

\begin{figure*}
\includegraphics{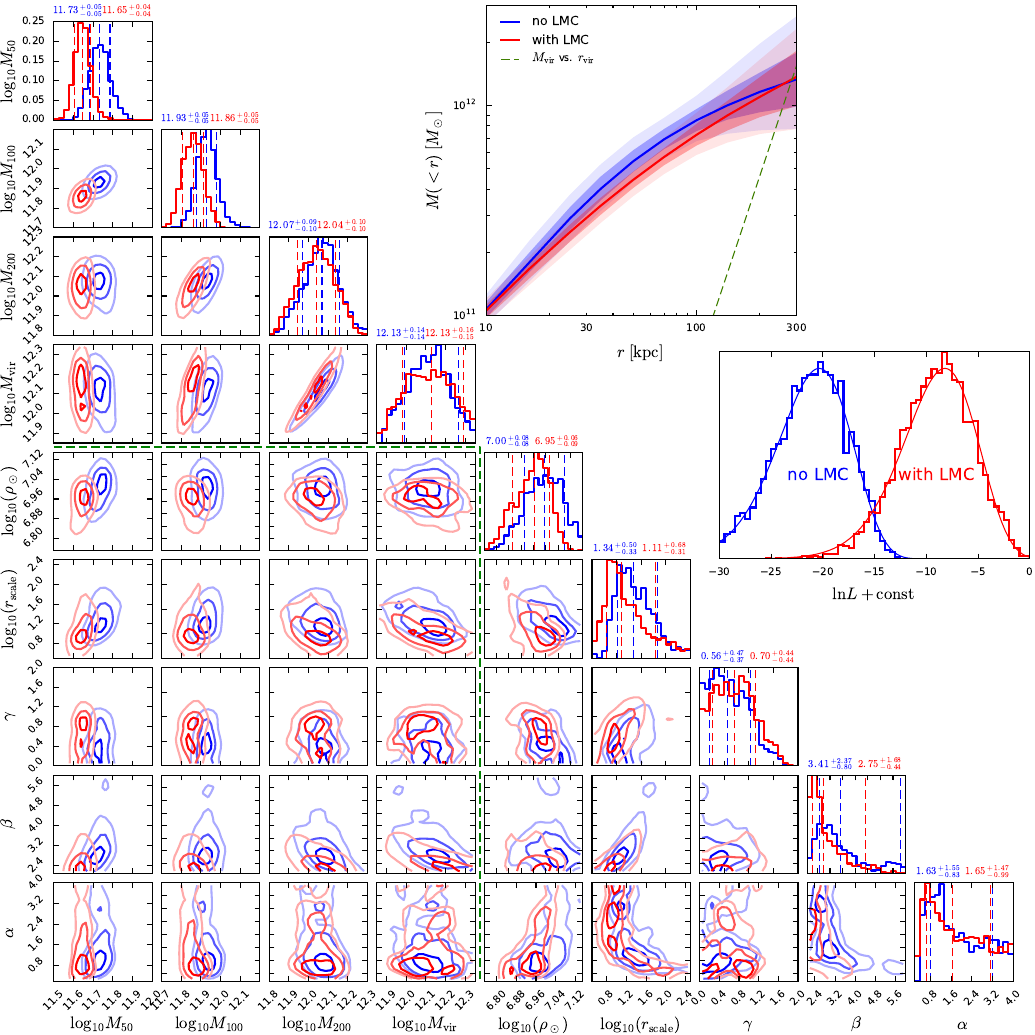}
\caption{Covariances and marginalized posterior distributions of potential parameters used in the fit (panels on the right of the green line) and derived enclosed masses at different radii (50, 100, 200 kpc and $r_\text{vir}$, panels on the left) for the Zhao family of halo profiles. We used the density $\rho_\odot$ at the Solar radius $r_\odot\equiv 8.2$~kpc instead of the normalization factor $\rho_0$ in the fit, since it is less strongly correlated with $r_\text{scale}$ and other parameters than $\rho_0$. Blue histograms and contours show the models without the LMC, and red -- with the LMC perturbation and orbit rewinding. The upper inset panel shows the enclosed mass profiles for both series of models: solid lines, darker and lighter shaded regions correspond to the median, 16/84 and 2.3/97.7 percentiles, and the green dashed line shows the relation between virial mass and radius. The right inset panel shows the histograms of likelihood values in both series of models, offset horizontally by a constant value (same for all models); overplotted thinner curves show the expected $\chi^2$ distributions with 28 degrees of freedom (the overall number of free parameters).
}  \label{fig:covplot_zhao}
\end{figure*}

\begin{figure*}
\includegraphics{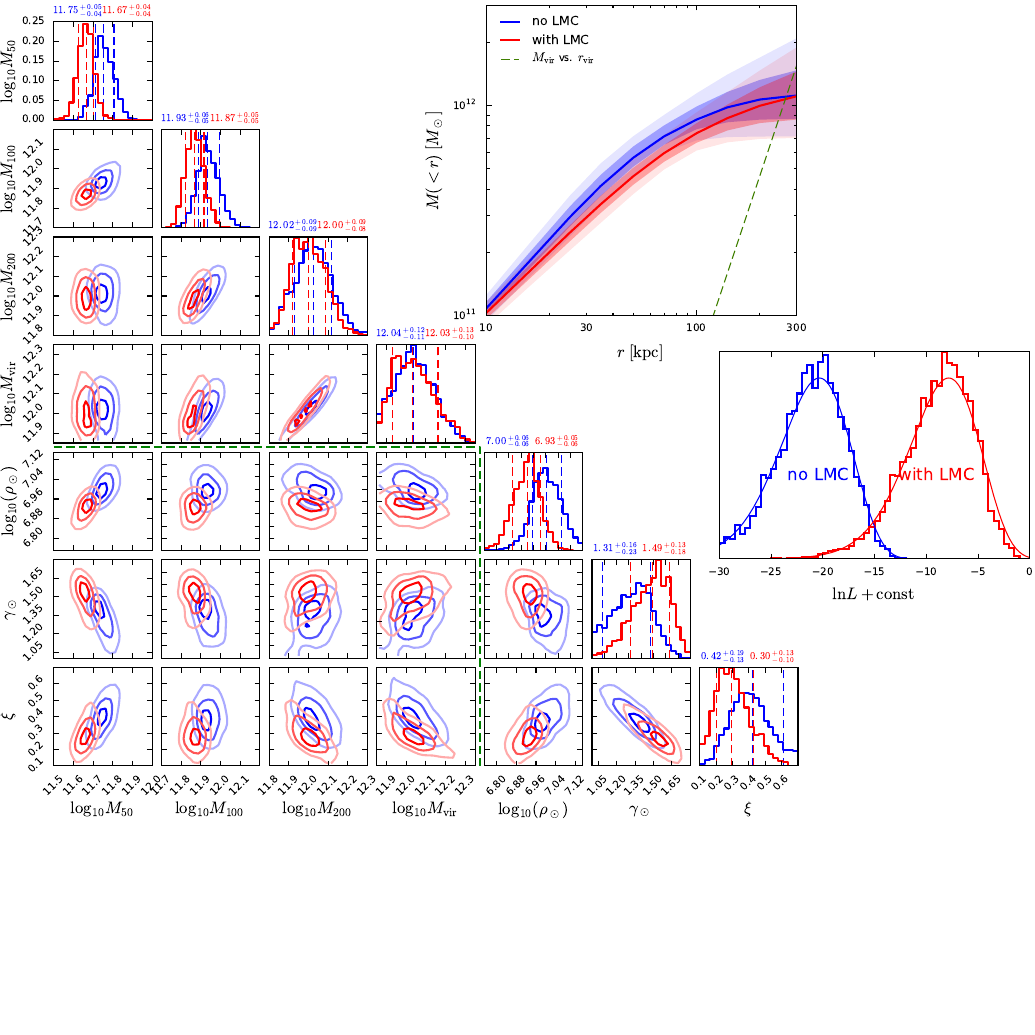}
\caption{Same as the previous figure, but for the Einasto family of models. Similarly to the Zhao family, we replace the intrinsic profile parameters $\rho_0$ and $r_\text{cut}$ by the equivalent pair $\rho_\odot$ (density at the Solar radius) and $\gamma_\odot$ (corresponding logarithmic density slope), since they are much less correlated than the intrinsic parameters. The enclosed mass profiles are very similar to the Zhao models for $r\lesssim 200$~kpc, but the density drops more steeply at larger radii, and therefore the extrapolated virial mass is somewhat lower (but with a large uncertainty in either case). The distribution of likelihoods is almost identical between both series of models (i.e., neither is statistically more preferred, although the Einasto family has fewer free potential parameters -- 3 instead of 5).
}  \label{fig:covplot_einasto}
\end{figure*}

\end{document}